\documentclass[onecolumn,notitlepage,aps,pra,10pt,amsmath,amssymb,showkeys,floatfix]{revtex4-2}
\usepackage{graphicx}
\usepackage{amsmath,amssymb,amsthm,amsfonts,bm}
\usepackage{dsfont}
\usepackage{xfrac}
\usepackage{booktabs}

\begin{document}

\title[The Rodeo Algorithm as a Quantum Kernel Method]{A Quantum Kernel Method for Density-of-States Estimation}

\author{Julio C. S. Rocha}
  \email{jcsrocha@ufop.edu.br}
  \affiliation{Departamento de Física, Universidade Federal de Ouro Preto - UFOP, Ouro Preto, Minas Gerais, Brasil.}

\begin{abstract}
We introduce a quantum kernel method for estimating the density of states (DoS) of quantum many-body systems. The method exploits the Rodeo algorithm as a physical implementation, requiring no modification of its standard single-ancilla circuit. Averaging the Rodeo response over Haar-random input states reconstructs the DoS convolved with a spectral kernel whose analytical form is determined entirely by the probability distribution of the sampled evolution times. This establishes a direct quantum analogue of the classical Kernel Polynomial Method (KPM): Haar-random states perform stochastic trace estimation, while the temporal sampling distribution plays the role of the reconstruction kernel. We derive the estimator, quantify its statistical and systematic uncertainties, and establish an explicit correspondence between quantum reconstruction kernels and classical signal-processing window functions, enabling concepts such as spectral leakage, side-lobe attenuation, and main-lobe width to guide kernel design. The method is validated numerically on the one-dimensional transverse-field spin--$\sfrac{1}{2}$ and spin--$1$ Ising models, where the reconstructed density of states and the resulting thermodynamic observables agree with exact diagonalization within the estimated statistical uncertainties.
\end{abstract}

  % keywords can be removed
\keywords{Quantum Computing, Kernel Method, Density of States, Transverse Field Ising Model, Qudit}
\maketitle
%%%%%%%%%%%%%%%%%%%%%%%%%%%%%%%%%%%%%%%%%%%%%%%%%%%%%%%%%%%%%%%%%%%%%%%%%%%%%%%
\section{Introduction}
\label{sec:intro}

The density of states (DoS) is a fundamental quantity in many-body physics. It provides the basis for calculating thermodynamic properties through the microcanonical entropy~\cite{Goold2016,gross2001,Rocha2025}, contributes to the characterization of transport and linear-response functions~\cite{Kubo1957-lw,mahan2000,ashcroft1976}, and provides information about the spectral signatures of quantum and classical phase transitions~\cite{gross2001,qi2018}. Efficient methods for estimating the DoS are therefore central to computational physics.

Among classical approaches for matrices too large to be diagonalized exactly, stochastic trace-based methods provide a particularly useful framework for estimating the DoS. These methods combine a stochastic approximation to the trace with a spectral reconstruction procedure. The kernel polynomial method (KPM)~\cite{weisse2006,silver1994} is a standard example. In KPM, the DoS is expanded in Chebyshev polynomials whose moments are estimated stochastically, and the truncated expansion is multiplied by a damping kernel, typically the Jackson kernel, to suppress Gibbs oscillations arising from the finite expansion order. An analogous strategy can be formulated in the time domain. In the equation-of-motion method introduced by Alben \textit{et al.}~\cite{Alben1975}, the time-dependent Schrödinger equation is integrated numerically for initial states with random phases, and the DoS is reconstructed from the Fourier transform of the resulting autocorrelation function. Off-diagonal contributions to the stochastic trace estimate decrease with the number of independent realizations, while the finite propagation time and the associated time window determine the spectral resolution. In this sense, the time window plays a role analogous to the reconstruction kernel in KPM.

Hams and De~Raedt~\cite{Hams2000} subsequently provided a systematic analysis of the time-domain approach, including the statistical errors associated with stochastic trace estimation. Under suitable assumptions, the statistical error decreases with both the number of independent random vectors and the Hilbert-space dimension, explaining why relatively few random vectors can suffice for estimating global spectral quantities in large systems. They also analyzed errors arising from numerical time propagation. Nevertheless, both KPM and time-domain methods require operations on vectors of dimension \(N_s\), with computational costs typically scaling linearly in \(N_s\) for sparse, bounded-connectivity Hamiltonians. Since \(N_s=d^N\) for a system of \(N\) constituents with local Hilbert-space dimension \(d\), these methods ultimately face an exponential scaling of memory and computational resources with system size.

This exponential growth poses a fundamental obstacle to classical simulations of general quantum many-body systems, as discussed by Abrams and Lloyd~\cite{abrams1997}. Quantum computers offer a potential way to overcome this representation bottleneck by encoding and manipulating quantum states directly in quantum registers~\cite{feynman1982,lloyd1996}. For suitable local Hamiltonians, their time evolution can be approximated using product formulas implemented as quantum circuits. This makes time-domain spectral estimation a natural candidate for quantum implementation. Indeed, De~Raedt \textit{et al.}~\cite{DeRaedt2000} reformulated the equation-of-motion method as a quantum algorithm for many-body spin systems and tested it using a software emulation of a quantum computer with up to 21 qubits. On physical quantum hardware, however, the autocorrelation \(\langle \psi | e^{-iHt} | \psi \rangle\) is not generally obtained through a direct measurement and must instead be accessed interferometrically, for example through ancilla-controlled time evolution.

Several quantum algorithms for spectral density estimation have since been developed, employing complementary strategies based on polynomial expansions, quantum phase estimation, and time-domain signal processing~\cite{abrams1999,roggero2020,somma2019,obrien2019,lin2022}. Summer \textit{et al.}~\cite{summer2024} implemented a quantum algorithm inspired by KPM, combining stochastic trace estimation with controlled quantum operations to evaluate Chebyshev moments. They demonstrated the approach on a trapped-ion quantum processor for a nonintegrable spin chain with up to 18 qubits. More recently, Goh and Koczor~\cite{goh2026} developed a quantum algorithm to estimate the Fourier transform of the DoS from Loschmidt amplitudes, measured via Hadamard tests with a single ancilla qubit. They showed that random initial states generated by unitary \(1\)-designs, including computational-basis states sampled uniformly through random bit flips, suffice to obtain an unbiased estimator of the Fourier transform of the DoS. This reduces the state-preparation requirements of earlier approaches based on \(2\)- and \(3\)-designs~\cite{Dankert2009}. Their method also extends to fixed-particle-number subspaces, enabling the estimation of subspace-resolved DoS relevant to fermionic systems. Because the accessible evolution time is finite, the reconstructed DoS is convolved with a Gaussian window, whose width determines the spectral resolution. Their numerical results further demonstrate that low-resolution reconstructions can be robust against Trotter errors and gate noise. Together, these works illustrate complementary approaches to quantum DoS estimation based on polynomial expansions and time-domain reconstruction.

The Rodeo algorithm~\cite{choi2021rodeo,qian2024demonstration,Gomes2025} provides an alternative framework for ancilla-based spectral filtering. We previously proposed a method for estimating the DoS of many-body quantum systems using the Rodeo algorithm~\cite{Rocha2024}. In this first protocol, contributions from individual computational-basis input states are accumulated, requiring an exponentially growing number of distinct input-state sweeps as the system size increases. We subsequently introduced a protocol that avoids this explicit summation for systems whose energy eigenstates coincide with the computational basis states~\cite{rocha2026qudit}. By preparing a homogeneous superposition of computational-basis states, which assigns uniform spectral weights in this special case, the protocol enables DoS estimation through a single sweep over the target energy. Its applicability, however, is restricted by the requirement that the computational basis coincide with the energy eigenbasis.

In this work, we extend these developments and introduce a quantum kernel method for estimating the DoS of general many-body quantum systems without explicitly sweeping over every computational-basis input state. We show that averaging the Rodeo response over Haar-random input states yields an estimator proportional to the DoS convolved with a spectral kernel determined by the Rodeo response and the probability distribution of the sampled evolution times. Unlike methods that reconstruct the DoS from the Fourier transform of time-domain autocorrelation functions, our approach obtains a spectrally filtered estimate directly from the energy-dependent Rodeo response. The resulting formulation connects Rodeo-based spectral filtering with the general structure of stochastic trace estimation and kernel-based spectral reconstruction. It provides an analytical framework for characterizing the smoothing and spectral-resolution properties of the estimator, extending the application of the Rodeo algorithm beyond eigenvalue localization.

The remainder of this paper is organized as follows. In
Sec.~\ref{sec:method}, we derive the proposed estimator.
Section~\ref{sec:response} reviews the Rodeo algorithm, while
Sec.~\ref{sec:SA} develops its statistical properties and
Sec.~\ref{sec:typicality} derives the density-of-states estimator.
Section~\ref{sec:error} discusses the principal sources of uncertainty,
including the statistical properties of the estimator
(Sec.~\ref{sec:statError}) and the errors associated with the
Suzuki--Trotter decomposition and realistic quantum hardware
(Sec.~\ref{sec:trotter}). In Sec.~\ref{sec:dictionary}, we establish the
correspondence between the proposed quantum kernel method and the
classical kernel polynomial method. Numerical demonstrations for the
one-dimensional spin--$\sfrac{1}{2}$ and spin--$1$ transverse-field Ising models are presented in Sec.~\ref{sec:results}. Finally,
Sec.~\ref{sec:discussion} summarizes the main conclusions and discusses
possible extensions of the proposed framework.

%%%%%%%%%%%%%%%%%%%%%%%%%%%%%%%%%%%%%%%%%%%%%%%%%%%%%%%%%%%%%%%%%%%%%%%%%%%%%%%
\section{Methodology}
\label{sec:method}
%%%%%%%%%%%%%%%%%%%%%%%%%%%%%%%%%%%%%%%%%%%%%%%%%%%%%%%%%%%%%%%%%%%%%%%%%%%%%%%
\subsection{Rodeo Response of an Arbitrary State: Rodeo Kernel}
\label{sec:response}

\begin{figure*}[h]
\begin{center}
\includegraphics[scale=0.15]{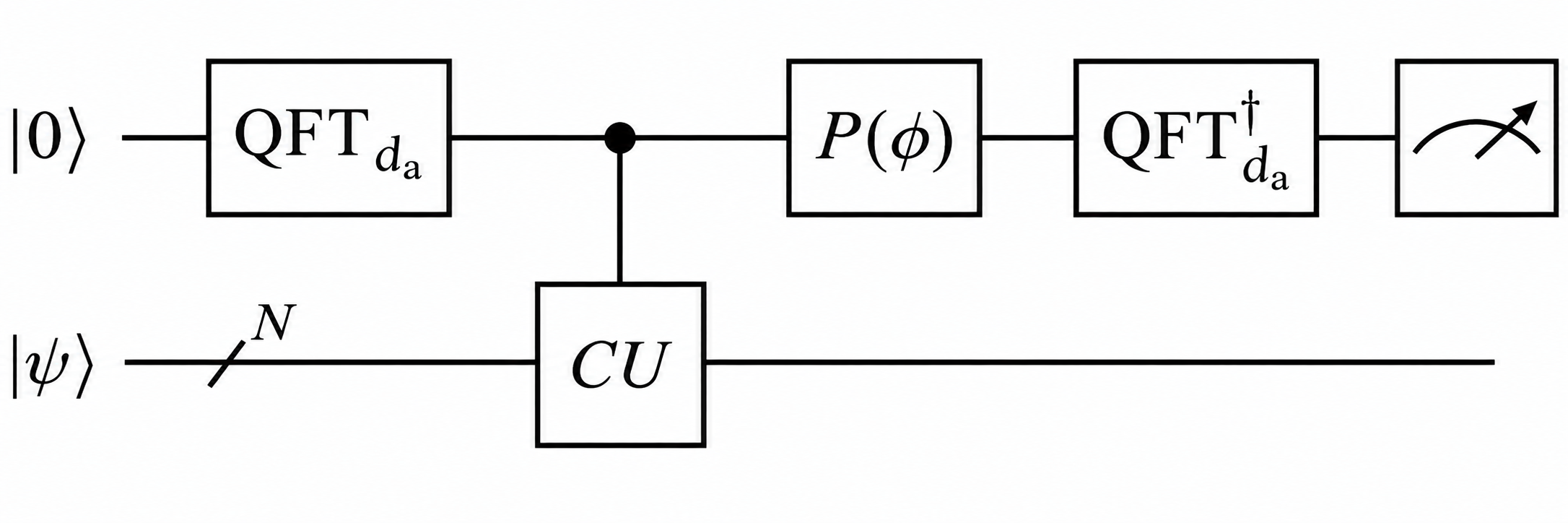}
\end{center}
\caption{Quantum circuit implementing a single Rodeo cycle with an ancilla
qudit of dimension $d_a$. The system register is initialized in an
arbitrary state $|\psi\rangle$, while the ancilla is prepared in the
computational basis state $|0\rangle$. A quantum Fourier transform (QFT)
prepares the ancilla in an equal superposition of its basis states, which
controls the time-evolution operator,
$CU$, for a randomly sampled evolution time $t$. A phase shift $P(\phi)$ is subsequently 
applied, followed by the inverse quantum Fourier transform ($\mathrm{QFT}^{\dagger}$). 
Finally, the ancilla is measured in the computational basis.}
\label{fig:RodeoCircuit}
\end{figure*}

We employ the qudit implementation of the Rodeo algorithm introduced in
Ref.~\cite{rocha2026qudit}. The complete circuit construction is described therein;
 here we summarize only the ingredients required to derive the DoS estimator. Qudits generalize qubits to quantum systems with \(d_a\) distinct levels.
 The quantum circuit implementing the Rodeo algorithm is illustrated in Fig.~\ref{fig:RodeoCircuit}.

The first step of the algorithm is to prepare the ancilla qudit in the
equal-superposition state, which we denote $|+\rangle$. This is achieved by 
applying the $d_a$-dimensional quantum Fourier transform ($\mathrm{QFT}_{d_a}$) to the
computational basis state $|0\rangle$, namely, $|+\rangle = \mathrm{QFT}_{d_a}|0\rangle$,
where
\begin{equation}
\mathrm{QFT}_{d_a}
=
\frac{1}{\sqrt{d_a}}
\sum_{\ell,n=0}^{d_a-1}
\exp{\left(\frac{i 2\pi \ell n}{d_a}\right)}
|\ell\rangle\langle n|.
\label{eq:QFT}
\end{equation}
Next, a controlled time-evolution operator,
\begin{equation}
CU
=
\sum_{n=0}^{d_a-1}
|n\rangle\langle n|
\otimes
\left(
e^{-i\mathcal{H}_s t}
\right)^n,
\end{equation}
is applied to the system register, where the ancilla qudit acts as the control. Subsequently, the ancilla undergoes a phase shift
\begin{equation}
P(E,t)
=
\sum_{n=0}^{d_a-1}
e^{inEt}
|n\rangle\langle n|,
\end{equation}
as an attempt to reverse the phase accumulated during the controlled time evolution. An inverse quantum Fourier transform, $\mathrm{QFT}_{d_a}^{\dagger}$, is then applied to map the ancilla back to the computational basis. Finally, the ancilla is measured in the computational basis, which is the measurement of the expectation value of the clock operator,
\begin{equation}
Z_{d_a}
=
\sum_{\ell=0}^{d_a-1}
e^{2\pi i\ell/d_a}
|\ell\rangle\langle\ell|.
\end{equation}

Let the system Hamiltonian satisfy
\begin{equation}
\mathcal H_s|E_k\rangle
=
E_k|E_k\rangle,
\end{equation}
and consider an arbitrary input state
\begin{equation}
|\psi\rangle
=
\sum_{k=1}^{N_s}
c_k|E_k\rangle.
\label{eq:psiExpansion}
\end{equation}
Here, $N_s=d_s^N$ is the dimension of the many-body Hilbert space, where
$d_s$ is the number of internal states of each particle and $N$ is the
total number of particles. Since the controlled time evolution is diagonal in the energy basis, each energy eigenstate evolves independently through the Rodeo circuit. Moreover,
the measured observable acts exclusively on the ancilla register,
$Z_{d_a}\otimes \mathds{1}_{s}$, where $\mathds{1}_{s}$ is the identity operator acting on the $N_s$-dimensional Hilbert space. Consequently, the off-diagonal coherences between
different energy eigenstates do not contribute to the expectation value,
and the Rodeo response depends only on the populations $|c_k|^2$. The expectation value of the ancilla qudit for a given probe energy, $E$, evolution time, $t$, initial state, $|\psi\rangle$, can be written as 
\begin{align}\nonumber
\mathcal{Z}(E,t,\psi) & := \langle Z_{d_a}\otimes \mathds{1}_{s}\rangle \\
& =
\sum_{k=1}^{N_s}
|c_k|^2
K_{d_a}(\Delta_k,t).
\label{eq:H_echo}
\end{align}
where
\begin{equation}
K_{d_a}(\Delta_k,t)
=
\frac{d_a-1}{d_a}e^{-i\Delta_k t}
+
\frac{1}{d_a}e^{i\Delta_k' t},
\label{eq:Kernel}
\end{equation}
is the Rodeo kernel, with $\Delta_k=E_k-E$ and
$\Delta_k'=(d_a-1)\Delta_k$. The qubit case ($d_a=2$) constitutes a special limit in which the two frequency components become identical, $\Delta_k'=\Delta_k$. The two complex exponentials therefore combine into a single real oscillation,
\begin{equation}
K_{2}(\Delta_k,t)
=
\cos(\Delta_k t).
\label{eq:Kernelqubit}
\end{equation}

By inspecting eqn.~(\ref{eq:Kernel}), one can see that the Rodeo kernel consists of the superposition of two oscillatory components: a dominant contribution with frequency $\Delta_k$ and amplitude $(d_a-1)/d_a$, and a secondary contribution with frequency $(d_a-1)\Delta_k$ and amplitude $1/d_a$. Their interference is responsible for the characteristic noise reduction and peak narrowing of the ancilla qudit implementation compared to the ancilla qubit implementation, with the strongest effect observed for the qutrit one ($d_a = 3$), see Appendix~\ref{app:quditD}. 

%%%%%%%%%%%%%%%%%%%%%%%%%%%%%%%%%%%%%%%%%%%%%%%%%%%%%%%%%%%%%%%%%%%%%%%%%%%%%%%%%%%%%%%%%%%%%%%%%%
\subsection{Ensemble-Average: Spectral Kernel}
\label{sec:SA}

If the evolution times are sampled from an arbitrary
probability distribution $p(t)$, it is convenient to introduce its
characteristic function,
\begin{equation}
\Phi(\omega)
=
\int_{-\infty}^{\infty}
p(t)e^{i\omega t}\,dt.
\label{eq:charfun}
\end{equation}

Averaging $\langle Z_{d_a}\otimes \mathds{1}_{s}\rangle$, eqn.~(\ref{eq:H_echo}), over the temporal distribution yields the \emph{Spectral Amplitude} (SA) as
\begin{equation}
\mathcal{R}_{d_a}(E,\psi)
=
\sum_{k=1}^{N_s}
|c_k|^2
G_{d_a}(\Delta_k),
\label{eq:mixture}
\end{equation}
where
\begin{equation}
G_{d_a}(\Delta_k)
=
\frac{d_a-1}{d_a}\Phi(-\Delta_k)
+
\frac{1}{d_a}\Phi(\Delta_k').
\label{eq:Ggeneral}
\end{equation}
Eqn.~(\ref{eq:mixture}) shows that the SA is a linear functional of the spectral weights $|c_k|^2$, with kernel $G_{d_a}(\Delta_k)$, which will be hereafter referred to as the spectral kernel. 

Originally, the evolution times are sampled from a normal (Gaussian) distribution, i.e.,
\begin{equation}
p(t)
=
\frac{1}{\sqrt{2\pi}\sigma}
\exp\!\left(
-\frac{(t-\mu)^2}{2\sigma^2}
\right),
\end{equation}
where $\mu$ and $\sigma$ stand for the mean and the
standard deviation of the distribution, respectively. 
The characteristic function for this distribution is
\begin{equation}
\Phi(\omega)
=
e^{-\frac{\sigma^2\omega^2}{2}} e^{i\mu \omega}.
\end{equation}
To eliminate the oscillatory phase factor in the spectral kernel and avoid artificial modulations that could be misinterpreted as spectral features, we set $\mu=0$ throughout this work. These considerations lead to
\begin{equation}
G_{d_a}(\Delta_k)
= 
\frac{d_a-1}{d_a} 
\exp{\left(-\frac{\sigma^2\Delta_k^2}{2}\right)} +
\frac{1}{d_a}
\exp{\left(-\frac{\sigma^2(d_a-1)^2\Delta_k^{2}}{2}\right)}.
\label{eq:SA_general}
\end{equation}
%
%\begin{widetext}
%\begin{equation}
%G_{d_a}(\Delta_k)
%=
%\frac{d_a-1}{d_a}
%\exp{\left(-\frac{\sigma^2\Delta_k^2}{2}\right)}
%+
%\frac{1}{d_a}
%\exp{\left(-\frac{\sigma^2(d_a-1)^2\Delta_k^{2}}{2}\right)}.
%\label{eq:SA_general}
%\end{equation}
%\end{widetext}
%
The spectral kernel is therefore composed of two Gaussian contributions. The dominant component, with weight $(d_a-1)/d_a$, has a characteristic width determined by $\sigma$. In contrast, the second component carries a smaller weight of $1/d_a$ and exhibits faster Gaussian decay due to the larger effective frequency $(d_a-1)\Delta_k$. Consequently, for $d_a>2$, the second term provides only a subleading correction to the SA, slightly reducing the width of the Gaussian peak. For the qubit implementation ($d_a=2$), the spectral kernel reduces to
\begin{equation}
G_{2}(\Delta_k)
=
e^{-\frac{\sigma^2}{2} \Delta_k^2}.
\label{eq:SA_qubitNormal}
\end{equation}
Eqn.~(\ref{eq:mixture}) immediately suggests a DoS
estimator: if the input state is chosen randomly, the spectral weights
become uniformly distributed on average over the Hilbert space. The
ensemble-averaged Rodeo response is therefore proportional to the DoS convolved with the spectral filter $G(\Delta_k)$, as shown in the
next subsection.

%%%%%%%%%%%%%%%%%%%%%%%%%%%%%%%%%%%%%%%%%%%%%%%%%%%%%%%%%%%%%%%%%%%%%%%%%%%%%%%
\subsection{The Estimator - Random states and typicality}
\label{sec:typicality}

The key idea is to sample the state $|\psi\rangle$ according to the Haar-uniform measure over the $N_s$-dimensional Hilbert space. Owing to the unitary invariance of the Haar measure and the normalization condition $\sum_k |c_k|^2 = 1$, each basis coefficient is statistically equivalent, implying
\begin{equation}
    \mathbb{E}\!\left(|c_k|^2\right)=\frac{1}{N_s}, \qquad \forall k.
\end{equation}
Consequently, taking the Haar average of eqn.~(\ref{eq:mixture}) for the spectral kernel given by eqn.~(\ref{eq:SA_qubitNormal}), yields
\begin{equation}
    \mathbb{E}(\mathcal{R}_{2}(E,\psi))
    =  \frac{1}{N_s} \sum_{k=1}^{N_s} e^{-\frac{\sigma^2}{2} (E-E_k)^2} 
    =  \frac{1}{N_s} \mathrm{Tr}\left[ e^{-\frac{\sigma^2}{2}  (E\mathds{1}_{s} - \mathcal{H}_s)^2} \right].
    \label{eq:estimator}
 \end{equation}
 Since,
 \begin{equation}
 (\Omega * G)(E) \equiv  \int \Omega(E')e^{-\frac{\sigma^2}{2}  (E-E')^2} \mathrm{d}E'
     = \mathrm{Tr}\left[ e^{-\frac{\sigma^2}{2}  (E\mathds{1}_s - \mathcal{H}_s)^2} \right],
\label{eq:estimatorB}
 \end{equation}
%\begin{widetext}
%\begin{equation}
 %   \mathbb{E}(\mathcal{R}_{2}(E,\psi))
 %   = \frac{1}{N_s} \sum_{k=1}^{N_s} e^{-\frac{\sigma^2}{2} (E-E_k)^2}
 %   =  \frac{1}{N_s} \mathrm{Tr}\left[ e^{-\frac{\sigma^2}{2}  (E\mathds{1}_{s} - \mathcal{H}_s)^2} \right],
 %   \label{eq:estimator}
 %\end{equation}
% Since,
%\begin{equation}
 %(\Omega * G)(E) \equiv \int \Omega(E')e^{-\frac{\sigma^2}{2}  (E-E')^2} \mathrm{d}E'  = \mathrm{Tr}\left[ e^{-\frac{\sigma^2}{2}  (E\mathds{1}_s - \mathcal{H}_s)^2} \right],
%\label{eq:estimatorB}
% \end{equation}
%\end{widetext}
one can say that the Haar-averaged response is the DoS ($g(E) = \Omega(E)/N_s$) convolved with the Gaussian~\cite{rocha2026qudit}. It should be emphasized that the additional Gaussian component of the spectral kernel in the qudit implementation introduces only a small correction to the convolution kernel.
The estimator from $R$ independent random states
$\psi_1,\dots,\psi_R$ is then
\begin{equation}
    \hat{g}(E)
    = \frac{1}{R} \sum_{r=1}^{R} \mathcal{R}_{d_a}(E,\psi_r),
    \label{eq:ghat}
\end{equation}
which is unbiased for $(g*G)$. 

At first, if the energy-level gap exceeds the width of the spectral kernel, individual eigenvalues give rise to well-resolved peaks, and their degeneracies can be inferred directly from each peak height. Otherwise, neighboring Gaussian peaks overlap and merge into a single broader feature, making the individual degeneracies impossible to distinguish. In principle, the spectral resolution can be improved arbitrarily by increasing the parameter $\sigma$, thereby narrowing the spectral kernel. However, as discussed in the next section, the corresponding increase in the evolution time makes such a refinement impractical due to the increasing cost and error associated with the Suzuki--Trotter decomposition.

Since only entropy differences are physically meaningful, an overall multiplicative constant in the DoS does not affect the thermodynamic properties of the system. We therefore interpret
$\hat{g}(E)\,\delta E$ as an unnormalized estimator of the DoS within the energy interval $[E,E+\delta E]$. Consequently, the degeneracy associated with each resolved energy level can be obtained by integrating $\hat{g}(E)$ over the corresponding peak. A normalization constant may be obtained by integrating the estimator over the entire energy spectrum.

%%%%%%%%%%%%%%%%%%%%%%%%%%%%%%%%%%%%%%%%%%%%%%%%%%%%%%%%%%%%%%%%%%%%%%%%%%%%%%%

\subsection{Sources of Uncertainty}
\label{sec:error}

In this section, we analyze the principal sources of uncertainty affecting
the reconstructed DoS. These can be broadly classified into
two categories: statistical uncertainties, arising from the stochastic
nature of the estimator, and implementation errors associated with the
quantum simulation of the time-evolution operator.

\subsubsection{Statistical Uncertainties}
\label{sec:statError}

The first source of statistical uncertainty arises from the use of Haar-random input states. For these states, the spectral weights $|c_k|^2$ follow a symmetric Dirichlet distribution with mean $1/N_s$~\cite{Zyczkowski2001}. In the large-Hilbert-space limit, the individual weights approach a Porter--Thomas (exponential) distribution~\cite{Porter1956}, with relative fluctuations of order unity; that is, their standard deviation is comparable to their mean.

Quantum typicality of an observable refers to the phenomenon whereby its expectation values in individual Haar-random states concentrate around the corresponding Haar average~\cite{Facchi2015}. In the present case, the reconstructed DoS is obtained from the Haar average of the SA, which corresponds to the DoS convolved with the spectral kernel, as shown in eqns.~(\ref{eq:estimator}) and (\ref{eq:estimatorB}). As shown in Appendix~\ref{app:typicality}, the standard deviation of the SA across Haar-random input states scales, in the large-Hilbert-space limit, as
\begin{equation}
    \sigma_{\mathrm{haar}}(\mathcal{R})
    = \mathcal{O}(N_s^{-1/2}).
\end{equation}
When the probe energy $E$ is far from all energy eigenvalues, the spectral kernel values are approximately zero, and hence $\sigma_{\mathrm{haar}}(\mathcal{R})\approx 0$.

In practice, the DoS estimator is obtained by averaging the SA over a finite number $R$ of independent Haar-random input states, as defined in eqn.~(\ref{eq:ghat}). The corresponding standard error due to Haar-state sampling is
\begin{equation}
    \sigma_{\mathrm{haar}}(\hat g)
    = \frac{\sigma_{\mathrm{haar}}(\mathcal{R})}{\sqrt{R}}
    = \mathcal{O}\left(\frac{1}{\sqrt{RN_s}}\right).
    \label{eq:sigmaHaar_g}
\end{equation}

A second source of statistical uncertainty arises from the finite number of evolution times sampled in each Rodeo sweep. The proposed protocol involves two distinct averaging procedures. The first is the quantum expectation value of the ancilla clock operator, discussed in the next section. The second is the ensemble average over randomly sampled evolution times, which we consider here.

For each initial state $|\psi_r\rangle$, the quantity $\mathcal{R}_{d_a}(E,\psi_r)$ in eqn.~(\ref{eq:ghat}) is estimated by taking the Gaussian-weighted average of the expectation value $\langle Z_{d_a}\otimes \mathds{1}_s\rangle$, defined in eqn.~(\ref{eq:H_echo}), over $N_t$ sampled evolution times, as described in Sec.~\ref{sec:SA}. As shown in Ref.~\cite{rocha2026qudit}, the standard error associated with this temporal sampling scales as
\begin{equation}
    \sigma_{\mathrm{gauss}}(\mathcal{Z})
    = \mathcal{O}(N_t^{-1/2}).
    \label{eq:sigmaGauss}
\end{equation}
Furthermore, $\sigma_{\mathrm{gauss}}(\mathcal{Z})=0$ when $|\psi_r\rangle$ is an eigenstate of the Hamiltonian with eigenvalue $E$.

Since the evolution times are sampled independently for each initial state, the finite-time estimates $\mathcal{R}_{d_a}(E,\psi_r)$ are independent across initial states. Their fluctuations reflect both Haar-state variability and temporal-sampling uncertainty. Assuming these contributions are independent, the variance of the estimator is therefore
\begin{equation}
    \sigma^2(\hat g)=\frac{1}{R}\Bigl[\sigma^2_{\mathrm{haar}}(\mathcal{R})
    +\langle\sigma^2_{\mathrm{gauss}}(\mathcal{Z})\rangle_r\Bigr],
\end{equation}
where $\langle\cdot\rangle_r$ denotes the average over the $R$ initial states.

In our numerical implementation, we estimate $\sigma(\hat g)$ directly from the finite-time SA estimates across the $R$ initial states, using their sample standard deviation:
\begin{equation}
    \sigma(\hat g)=\frac{s_{\mathcal{R}}}{\sqrt{R}},
    \label{eq:errorF}
\end{equation}
where $s_{\mathcal{R}}$ is the sample standard deviation of these estimates. This procedure incorporates both sources of uncertainty, so no separate temporal contribution is added. For $R=1$, $s_{\mathcal{R}}$ cannot be estimated; in this case, the error bar accounts only for temporal-sampling uncertainty (see Appendix~\ref{app:error}). Finally, even when the absolute standard deviation is small, the relative uncertainty, $\sigma(\hat g)/\hat g$, can remain large when the estimated DoS is small.

\subsubsection{Implementation Errors}
\label{sec:errRealMachine}

In an implementation on a physical quantum computer, the expectation value in eqn.~(\ref{eq:H_echo}) would be estimated by repeatedly executing and measuring the circuit over many repetitions (shots). Here $t$, $E$, and $|\psi_r\rangle$ are kept constant. This finite-shot sampling introduces an additional source of statistical uncertainty, $\sigma_{shots}(\mathcal{Z})$. We denote its contribution to the standard deviation of the DoS estimator by 
\begin{equation*}
 \langle\langle {\sigma^2_{shots}(\mathcal{Z})\rangle_t\rangle_r}  = \frac{1}{RN_t}\sum_{r=1}^R\sum_{t=1}^{N_t} \sigma^2_{shots}(\mathcal{Z}),
\end{equation*}
where the double brackets indicate the average of $\sigma^2_{shots}(\mathcal{Z})$ over both the sampled evolution times and the random initial states. 
Assuming independent measurement noise, this contribution adds to the variance associated with Haar-state and temporal sampling. 

\subsubsection{Trotter-error}
\label{sec:trotter}
Besides the statistical uncertainties discussed above, the quantum
implementation introduces additional systematic errors. The most important
of these is the Suzuki--Trotter decomposition employed to simulate the time evolution generated by Hamiltonians containing noncommuting terms.

Let the Hamiltonian be decomposed as
\begin{equation}
\mathcal H
=
\sum_{k=1}^{K}
\mathcal H_k,
\end{equation}
where the individual terms generally do not commute. The first-order
Suzuki formula is defined by
\begin{equation}
\mathcal{S}_1(t) =
e^{-i\mathcal H_1t}
\cdots
e^{-i\mathcal H_K t}.
\end{equation}

Following the simulation method of Lloyd~\cite{lloyd1996}, the error from a first-order Trotter formula can be written as
\begin{equation}
 \delta = \Big\| e^{-iHt}-\big[\mathcal{S}_1(t/r)\big]^r \Big\|
  \le \frac{t^2}{2r}\sum_{j=1}^{K}\sum_{k=j+1}^{K}\big\|[H_k,H_j]\big\|,
  \label{eq:S1bound}
\end{equation}
where $r$ is the number of Trotter steps~\cite{suzuki-91,hatano}. To achieve a
target precision $\delta$, we then consider the time dependence of the number of Trotter steps as
\begin{equation}
  r=\left\lceil \frac{t^2}{2\delta}\sum_{j<k}\big\|[H_k,H_j]\big\|\right\rceil.
  \label{eq:ntrotter}
\end{equation}
The parameter \(\delta\) specifies the target tolerance for the Suzuki–Trotter propagator approximation and is used to determine the number of Trotter steps. It is therefore a propagation-error tolerance, rather than a direct estimate of the error in the reconstructed DoS. The number of Trotter steps illustrates the trade-off between simulation accuracy and computational cost. Increasing either the Trotter order or the number of Trotter steps improves the approximation but simultaneously increases the quantum
resources required for the simulation~\cite{childs,Rocha2024}.

Moreover, implementing the algorithm with additional steps, or additional ancillary qudits, increases the spatial and temporal complexity of the circuit. Each ancilla requires state preparation, phase rotations, and controlled time-evolution operations, thereby increasing both the circuit width and its depth~\cite{barenco1995,bernstein1997,bouland2018}. Consequently, realistic implementations become more susceptible to gate imperfections and decoherence, which reduce the fidelity of the measured Rodeo response and therefore degrade the reconstructed DoS. As with any quantum algorithm based on coherent time evolution, fault-tolerant implementations will ultimately require quantum error-correction techniques to mitigate these effects~\cite{nielsen2010quantum,jaeger2006quantum,devitt}.
%%%%%%%%%%%%%%%%%%%%%%%%%%%%%%%%%%%%%%%%%%%%%%%%%%%%%%%%%%%%%%%%%%%%%%%%%%%%%%%

\subsection{The KPM dictionary}
\label{sec:dictionary}

The kernel formulation, introduced in Sec.~\ref{sec:SA}, establishes a close conceptual correspondence with the KPM~\cite{weisse2006}. Although the two approaches rely on different mathematical representations---Fourier reconstruction in the present work and Chebyshev polynomial expansions in KPM---both estimate the DoS by convolving the exact spectrum with a smoothing kernel. In both cases, the choice of kernel determines the spectral resolution, the suppression of oscillatory artifacts, and the overall reconstruction accuracy. This correspondence is summarized in Table~\ref{tab:analogy}.
\begin{table*}[h]
    \centering
    \caption{Conceptual correspondence between the classical KPM and the quantum
    kernel-based estimator.}
    \label{tab:analogy}
\begin{tabular}{l @{\hspace{0.5cm}} l}
\toprule
Classical KPM & Rodeo estimator \\
\midrule
random vectors (stochastic trace) & Haar-random initial states \\
Chebyshev expansion order $M$ & characteristic time scale of $p(t)$ \\
damping kernel & temporal sampling distribution $p(t)$ \\
Dirichlet kernel (no damping) & uniform distribution (sinc filter) \\
Gaussian (Silver--R\"oder) kernel & Gaussian distribution (Gaussian filter) \\
Jackson kernel & Hann (raised-cosine) sampling distribution \\
Gibbs oscillations & spectral leakage side lobes \\
\bottomrule
\end{tabular}
\end{table*}

The analogy extends beyond a simple comparison of reconstruction techniques.
The choice of $p(t)$ fully fixes the spectral
resolution, side-lobe structure, and leakage properties of the reconstructed
DoS. Rather than an implementation detail, $p(t)$ 
is thus the design parameter of the reconstruction kernel, given by eqn.~(\ref{eq:Ggeneral}).
Classical window functions developed for spectral estimation and signal processing 
therefore translate naturally into quantum reconstruction kernels with well-understood 
spectral properties. Representative examples are summarized in Table~\ref{tab:windows}.
\begin{table*}[h]
\centering
\caption{Temporal sampling distributions and the corresponding Rodeo
reconstruction kernels obtained through Fourier transformation.}
\label{tab:windows}
\small
\begin{tabular}{l l}
\toprule
Temporal sampling distribution $p(t)$ & Reconstruction kernel $G(E)$ \\
\midrule
Uniform            & $\mathrm{sinc}$ \\
Gaussian           & Gaussian \\
Hann (raised-cosine) & Linear combination of 3 shifted $\mathrm{sinc}$ functions \\
Hamming            & Weighted combination of 3 shifted $\mathrm{sinc}$ functions \\
Blackman           & Linear combination of 5 shifted $\mathrm{sinc}$ functions \\
Kaiser             & Bessel-type kernel \\
\bottomrule
\end{tabular}
\end{table*}

In both approaches the reconstruction is governed by a smoothing kernel whose Fourier or Chebyshev representation controls the trade-off between spectral resolution and oscillatory artifacts. For instance, a uniform sampling distribution over a finite interval,
\begin{equation}
p(t)=\frac{1}{2t_{\max}}, \qquad |t|\le t_{\max},
\end{equation}
produces a sinc kernel whose side lobes decay algebraically as $|E|^{-1}$ and alternate in sign. When convolved with a spectrum containing sharp features, these side lobes generate ringing analogous to the Gibbs phenomenon of an undamped Chebyshev expansion, and may even yield locally negative estimates of the otherwise non-negative DoS.

As in KPM, smooth kernel damping substantially suppresses these artifacts. The previously discussed zero-mean Gaussian sampling law eliminates the side lobes altogether, at the
cost of unbounded temporal support. As discussed in Sec.~\ref{sec:trotter}, increasing the evolution time $t$ requires a correspondingly larger Trotter number, which consequently increases the systematic error in the DoS reconstruction. Alternatively, the Hann sampling law,
%\begin{widetext}
\begin{equation}
p(t)\propto
\cos^2\!\left(\frac{\pi t}{2t_{\max}}\right),
\qquad |t|\le t_{\max},
\qquad
\langle t^2\rangle
=
t_{\max}^2\left(\frac13-\frac{2}{\pi^2}\right),
\end{equation}
%\end{widetext}
retains the experimentally convenient finite support while providing much
stronger leakage suppression. Because both $p(t)$ and its first derivative
vanish continuously at $|t|=t_{\max}$, its characteristic function decays
asymptotically as $|E|^{-3}$, compared with the $|E|^{-1}$ decay of the uniform
distribution, see Appendix~\ref{app:hann}.

This kernel-based perspective considerably broadens the scope of the Rodeo algorithm. Rather than restricting the analysis to Gaussian law, arbitrary temporal distributions may be employed to tailor the spectral reconstruction to specific experimental constraints or target resolutions. Consequently, the extensive body of knowledge on window design developed in signal processing becomes directly applicable to quantum DoS estimation. 

A systematic analysis of the reconstruction kernels associated with different temporal sampling distributions, together with their implications for spectral resolution, statistical efficiency, and robustness against decoherence, will be presented in a forthcoming publication.
%%%%%%%%%%%%%%%%%%%%%%%%%%%%%%%%%%%%%%%%%%%%%%%%%%%%%%%%%%%%%%%%%%%%%%%%%%%%%%%
\section{Results}
\label{sec:results}

In this section, we demonstrate the performance of the proposed scheme by
reconstructing the DoS using the estimator defined in
eqn.~(\ref{eq:ghat}). The spectral amplitude,
$\mathcal{R}_{d_a}(E,\psi)$, is evaluated using
eqn.~(\ref{eq:mixture}) together with the Gaussian spectral kernel of
eqn.~(\ref{eq:SA_general}), corresponding to a Gaussian sampling
distribution with zero mean ($\mu=0$). For each target energy, the Rodeo response is estimated by averaging over $N_t=1000$ independently sampled evolution times.

All calculations were carried out by numerically evolving the joint ancilla--system state throughout the Rodeo circuit using standard linear-algebra techniques. An ancilla qutrit ($d_a=3$) was employed in all simulations, as it has previously been shown to provide improved spectral resolution together with reduced statistical fluctuations compared with the qubit implementation~\cite{rocha2026qudit}, see Appendix~\ref{app:quditD}.

In all DoS reconstruction figures, black symbols represent the numerical estimates described above. Since our simulations evaluate the expectation values directly, they are free from shot noise. Consequently, the error estimate accounts only for statistical fluctuations arising from the finite sampling of evolution times and Haar-random initial states, as described in eqn.~(\ref{eq:errorF}). For $R=1$, the error bars account only for temporal-sampling uncertainty. Appendix~\ref{app:error} provides details of the data set, averaging procedures, and error-bar calculations. Vertical red lines represent the exact DoS obtained by direct Hamiltonian diagonalization.
\subsection{Spin Model}
\label{sec:model}
In addition to serving as a paradigmatic model in condensed matter
physics~\cite{sachdev2011}, the Ising Hamiltonian constitutes the standard representation of Quadratic Unconstrained Binary Optimization (QUBO) problems~\cite{lucas2014,kochenberger2014}. The addition of a transverse field introduces quantum fluctuations that form the basis of quantum
annealing~\cite{kadowaki1998,farhi2001}. Consequently, the transverse-field Ising model has become a benchmark for quantum optimization
algorithms~\cite{albash2018,hauke2020}. 

So, due to its importance, in this work, we consider a one-dimensional nearest-neighbor spin model described by the Hamiltonian
\begin{equation}
    \mathcal{H}_{d_s} =   -J\sum_{\langle i,j\rangle}S_i^z S_j^z - B \sum_i S^x_i,
\end{equation}
where $\langle i,j\rangle$ denotes nearest-neighbor pairs with $i=1,\ldots,N$ and $j=i+1$. Periodic boundary conditions are assumed, so that site $N+1$ is identified with site $1$. Here, $J$ denotes the exchange coupling constant, $B$ is the transverse-field coupling strength, and $S_{\ell}^{k}$ represents the spin operator along the $k$ direction acting on the $\ell$-th site of a local Hilbert space of dimension $d_s$. By varying $d_s$, this framework accommodates different spin models. In particular, the cases $d_s=2$ and $d_s=3$ correspond to the spin--$\sfrac{1}{2}$ and spin--$1$ transverse-field Ising models, respectively.  For notational simplicity, identity operators acting on all sites other than those explicitly indicated have been omitted. Throughout this work, energies are expressed in units of $JS^2$, where $S$ denotes the spin magnitude. Consequently, temperature is expressed in units of $JS^2/k_B$, and the transverse field is specified through the dimensionless ratio $B/(JS)$.

\subsection{One-Dimensional spin--$\sfrac{1}{2}$ Transverse Field Ising Model}
\label{sec:ising}

\begin{figure*}[h]
\begin{tabular}{c c}
\includegraphics[scale=0.28]{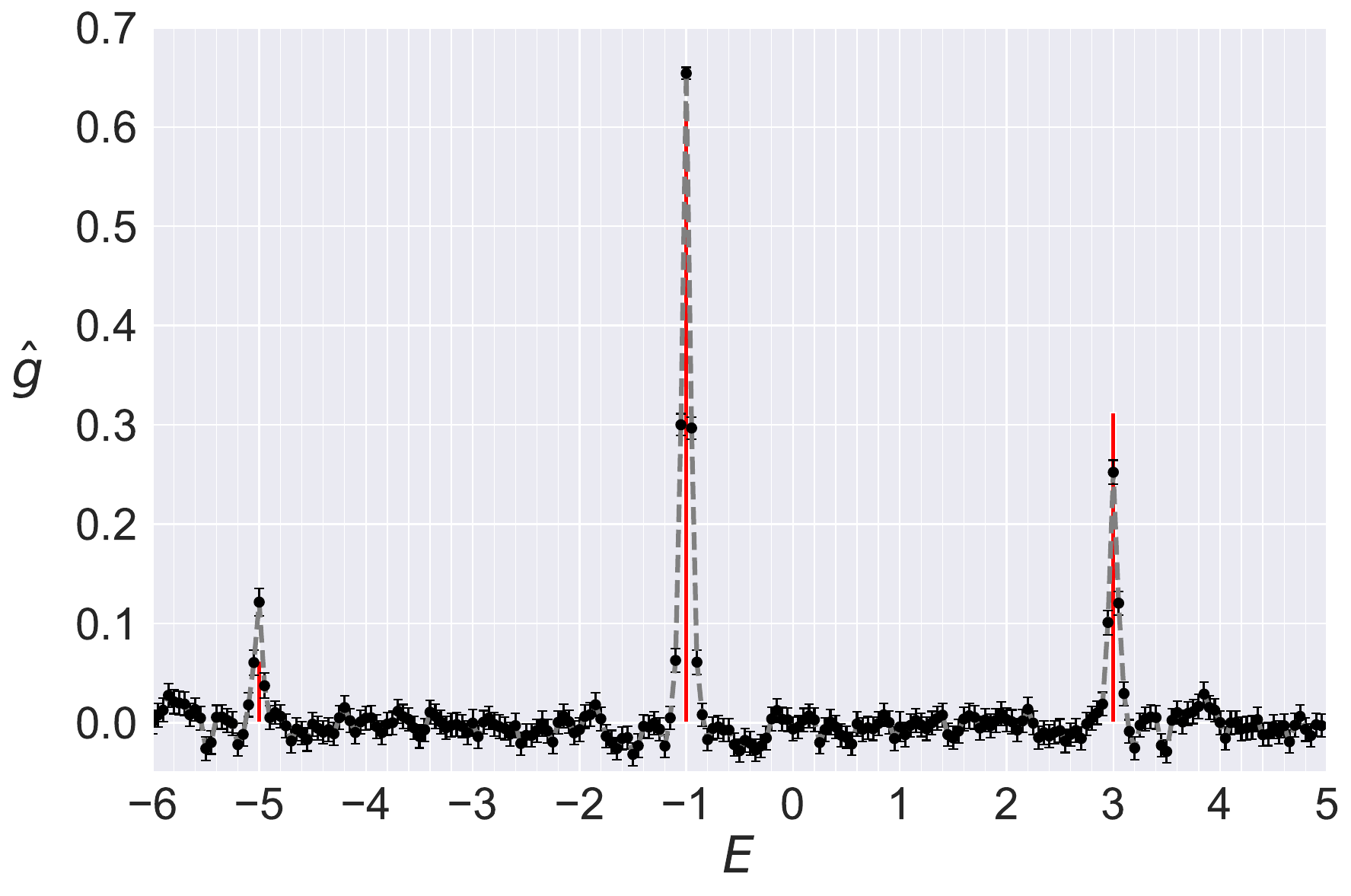} & 
\includegraphics[scale=0.28]{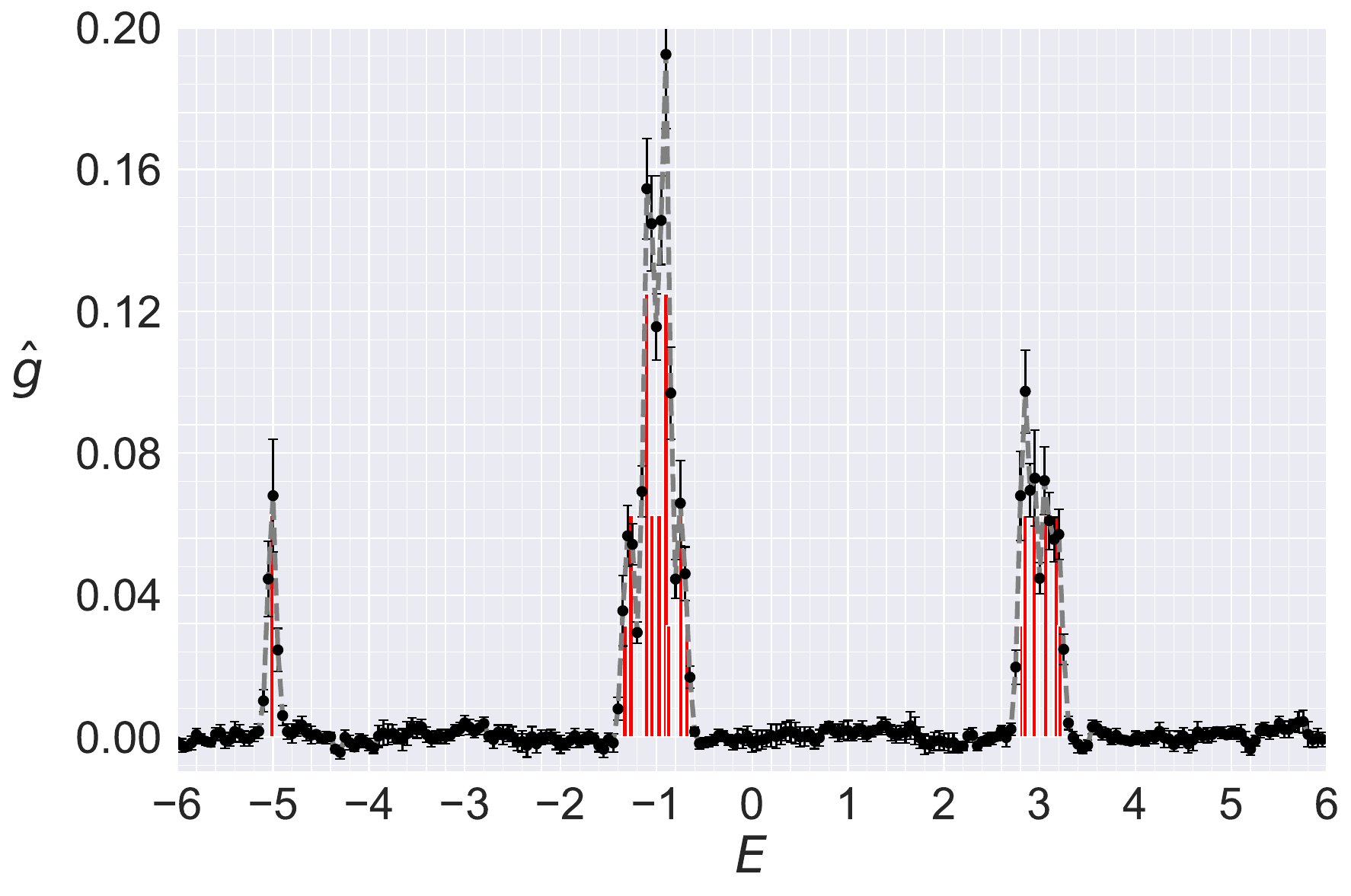} \\ (a) & (b) \\
\includegraphics[scale=0.28]{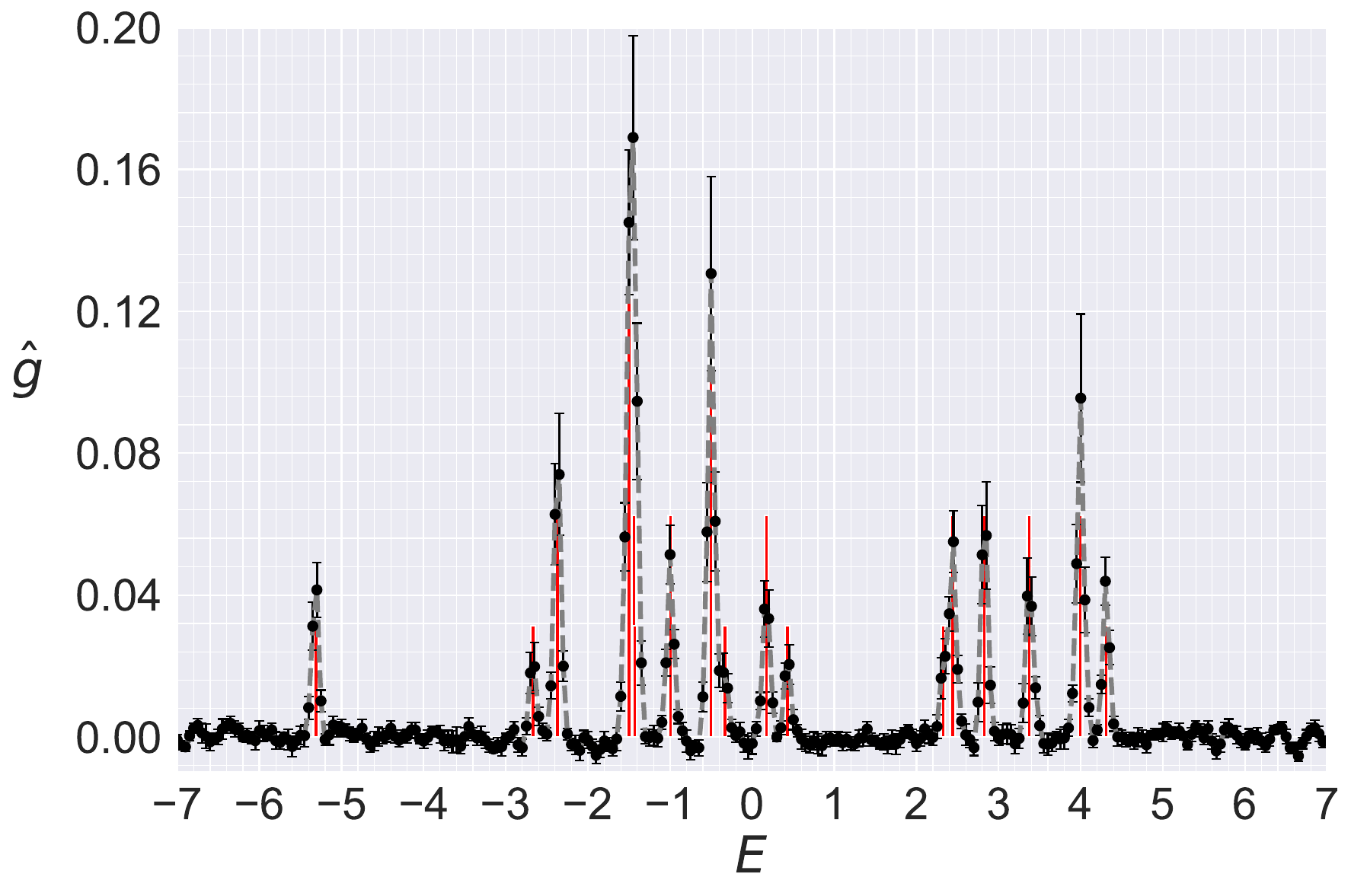} & 
\includegraphics[scale=0.28]{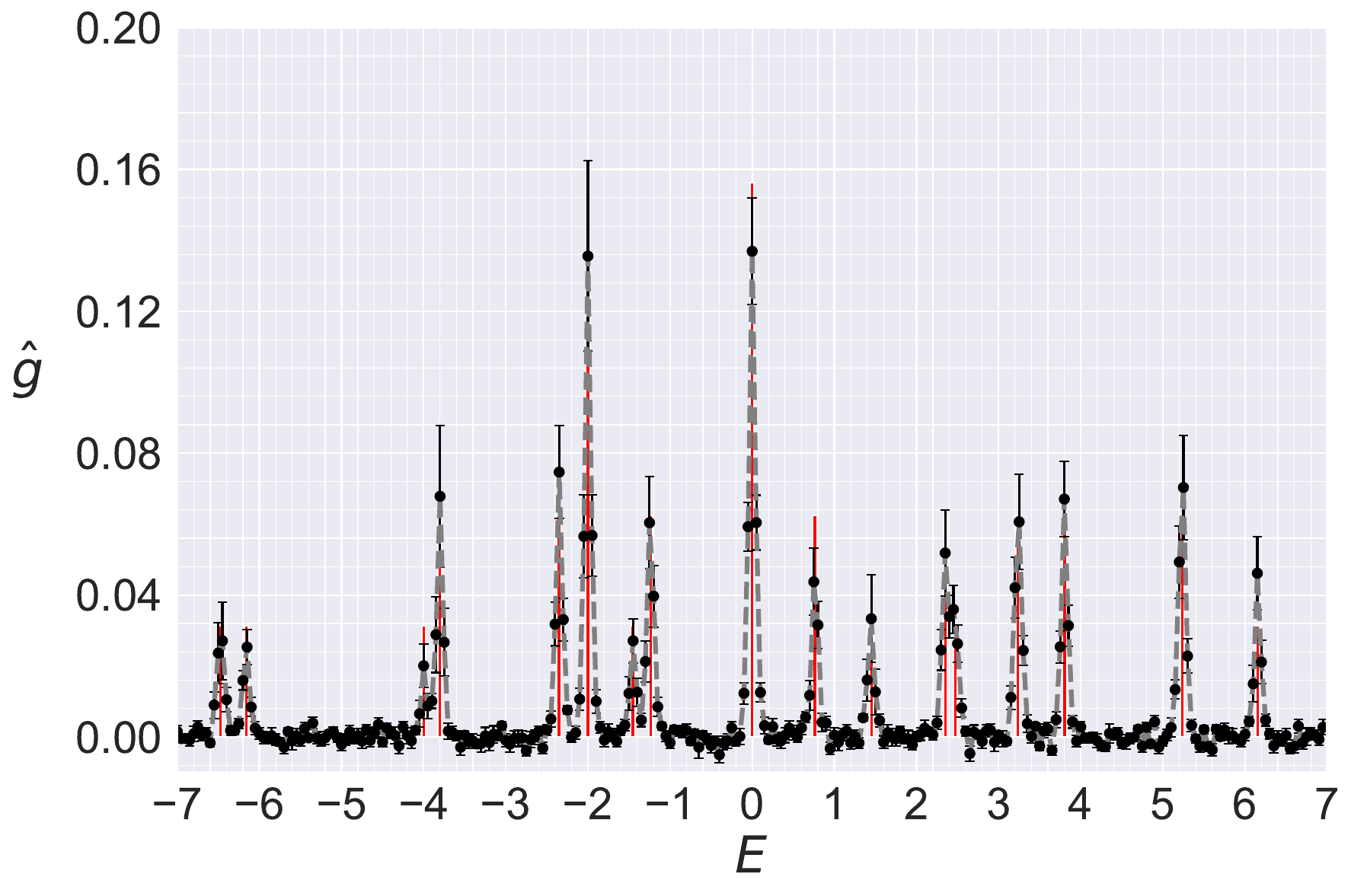} \\ (c) & (d)
\end{tabular}
\caption{Kernel-broadened DoS reconstruction for the 1D transverse-field Ising model using the Rodeo kernel method. For each target energy, the Rodeo algorithm was evaluated using $1\,000$ independent evolution times sampled from a Gaussian distribution with standard deviation $\sigma=20$. Panels (a)--(d) correspond to transverse-field strengths $B/(JS)=0, \ 0.1, \ 0.5$ and $1.0$, respectively. For $B\ne0$ the estimator was averaged over $R=10$ Haar-random initial states; only one state was used for $B=0.0$.}
\label{fig:isingHx}
\end{figure*}

In Fig.~\ref{fig:isingHx}, we present the reconstructed DoS for the Ising ring with $N=5$ particles in the presence of transverse-field strengths $B/(JS)=0$, $0.1$, $0.5$, and $1.0$, shown in panels (a)--(d), respectively. The reconstruction was performed using $R=10$ independent Haar-random input states, except for the case $B=0.0$, for which a single random state was employed.

In our previous work, we showed that a homogeneous superposition of the Hamiltonian eigenstates is sufficient to reconstruct the entire DoS from a single Rodeo energy sweep~\cite{rocha2026qudit}. For $B=0$, the Hamiltonian is diagonal in the computational basis, so its eigenstates coincide with the computational basis states. Moreover, the time-evolution operator can be evaluated exactly, and no Suzuki--Trotter decomposition is required.  In this case, we tentatively investigate whether a single Haar-random input state can reproduce the result obtained with the single-state microcanonical protocol proposed in our previous work. The corresponding numerical result is shown in Fig.~\ref{fig:isingHx}~(a). Comparison with the vertical red lines shows that the estimate is consistent with the value obtained by direct diagonalization.
However, in this situation, the single-input-state protocol is preferable because it requires fewer state preparations and yields more accurate results.

\begin{figure*}[h]
\begin{tabular}{c c}
\includegraphics[scale=0.28]{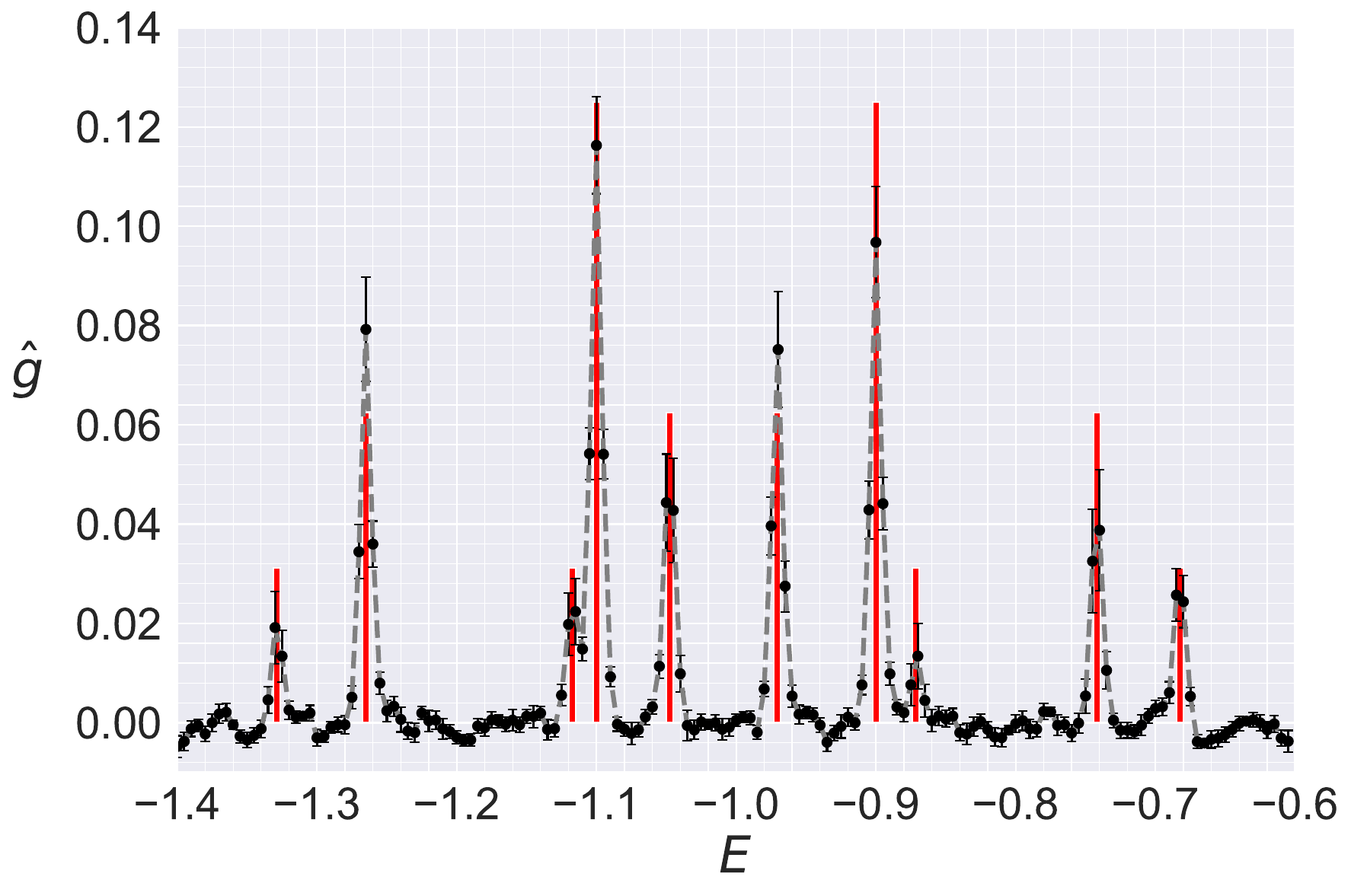} & 
\includegraphics[scale=0.28]{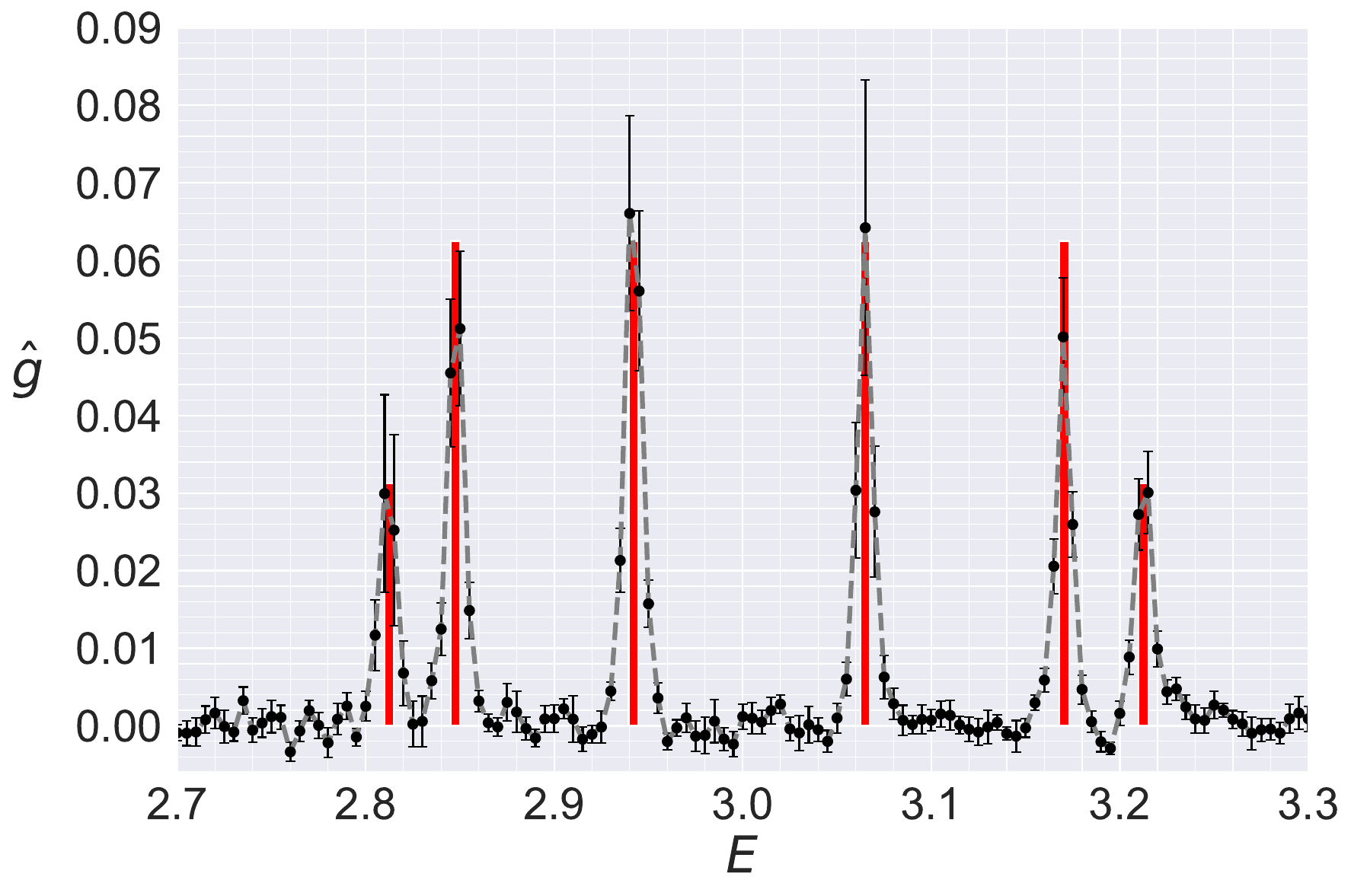} \\ (a) & (b) 
\end{tabular}
\caption{Refinement of the kernel-broadened DoS reconstruction for the
previously merged peaks shown in Fig.~\ref{fig:isingHx}(b) for the
transverse-field Ising model with $B/(JS)=0.1$. The estimator was averaged over
$R=10$ independent Haar-random initial states. For each target energy, the
Rodeo algorithm was performed using $1\,000$ evolution times sampled
independently from a Gaussian distribution with standard deviation
$\sigma=200$, yielding a narrower spectral kernel and improved energy
resolution. Panels (a) and (b) show enlarged views of the energy windows
$[-1.4,-0.6]$ and $[2.7,3.3]$, respectively.}
\label{fig:isingHxref}
\end{figure*}

The role of the magnetic field in lifting degeneracies and determining the energy splitting between the resulting levels is well known~\cite{zeeman,griffiths_introduction_2018}. This effect becomes progressively more pronounced as the magnetic field increases. For the relatively weak field $B/(JS)=0.1$, the highly degenerate energy levels present at $B=0.0$ broaden into clusters of closely spaced states, as shown in Fig.~\ref{fig:isingHx}~(b). For the parameters considered here, however, the energy separation between the split levels remains smaller than the width of the spectral kernel. Consequently, the individual contributions cannot be resolved and overlap to form a single broadened peak, yielding a smooth estimate of the DoS. The refinement of these broadened peaks is illustrated in Fig.~\ref{fig:isingHxref}, where the standard deviation of the Gaussian sampling distribution is increased to $\sigma=200$. This produces a narrower spectral kernel, thereby improving the energy resolution and allowing the previously merged peaks to be resolved. For both cases with $B/(JS)=0.1$, the Trotter error is $\delta\approx 0.03$. The corresponding mean numbers of Trotter steps are $\overline{r}=8{,}000$ for $\sigma=20$ and $\overline{r}=800{,}000$ for $\sigma=200$.

As the transverse magnetic field increases to $B/(JS)=0.5$, the splitting of the formerly degenerate energy levels becomes more pronounced and extends over a broader energy range [Fig.~\ref{fig:isingHx}~(c)]. Consequently, the reconstructed DoS becomes progressively flatter as the spectral weight is redistributed among a larger number of distinct energy levels. For $B/(JS)=1.0$ [Fig.~\ref{fig:isingHx}~(d)], well above the critical field $B_c/(JS)=0.5$~\cite{pfeuty}, this trend is further enhanced, yielding a considerably flatter DoS with smaller energy separations between neighboring levels. Although this evolution reflects the increasing influence of the transverse field and is qualitatively consistent with the quantum phase transition of the transverse-field Ising model, the present
results should not be interpreted as a determination of critical behavior\footnote{It is worth noting that the errors associated with the first-order Trotter approximation are $\delta\approx 0.15$ and $0.31$, for $B/(JS)=0.5$ and $1.0$, respectively.}.

For comparison, in Ref.~\cite{Rocha2024} the smoothed DoS was reconstructed by evaluating the Rodeo response for all $N_s=32$ computational basis states. Although that approach yields a smaller statistical error by eliminating fluctuations of the finite number of initial random states, it requires approximately three times more Rodeo sweeps than the present random-state implementation. Despite the good qualitative agreement between the two approaches, a direct quantitative comparison is not entirely appropriate, since the previous study employed only $500$ evolution times per energy and relied on an ancilla qubit, which intrinsically exhibits larger statistical fluctuations than the qutrit implementation.

\subsection{One-Dimensional Spin--$1$ Transverse Field Ising Model}
\label{sec:potts}

As an illustration of a system with a larger local Hilbert-space dimension, we consider the spin--$1$ model ($d_s=3$). The additional spin state increases the number of accessible magnetic configurations, making the DoS more sensitive to the applied transverse magnetic field. For the system size considered here ($N=5$), the Hilbert-space dimension is $N_s=3^5=243$. The reconstruction was performed using $R=36$ Haar-random input states. Because this number of input states is relatively small, the statistical uncertainty associated with quantum typicality constitutes the dominant source of fluctuations in the reconstructed DoS\footnote{The errors associated with the first-order Trotter approximation are $\delta\approx 0.007$ and $0.03$, for $B/(JS)=0.05$ and $0.2$, respectively.}.

\begin{figure*}[h]
\begin{tabular}{c c}
\includegraphics[scale=0.3]{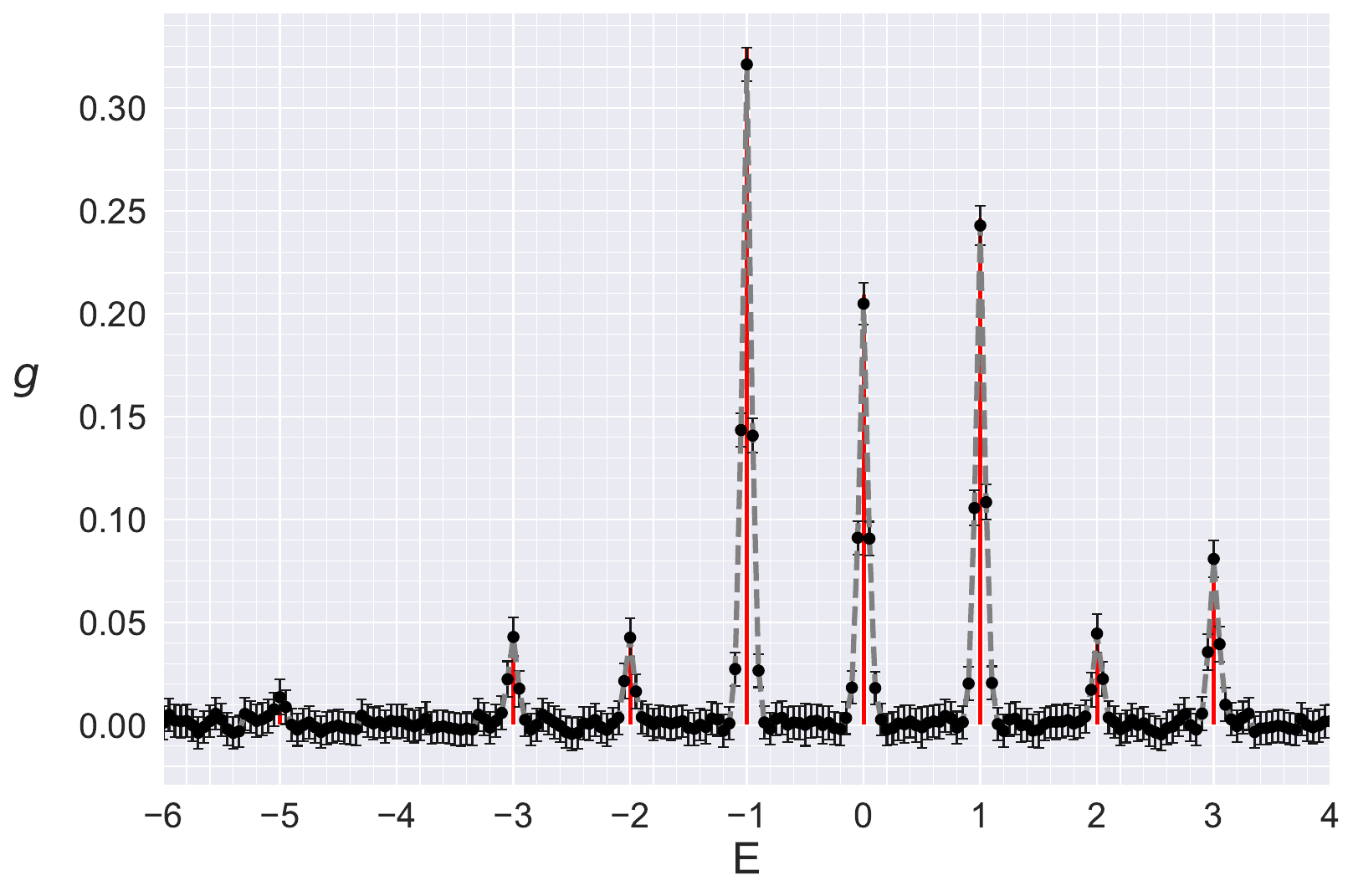} & 
\includegraphics[scale=0.3]{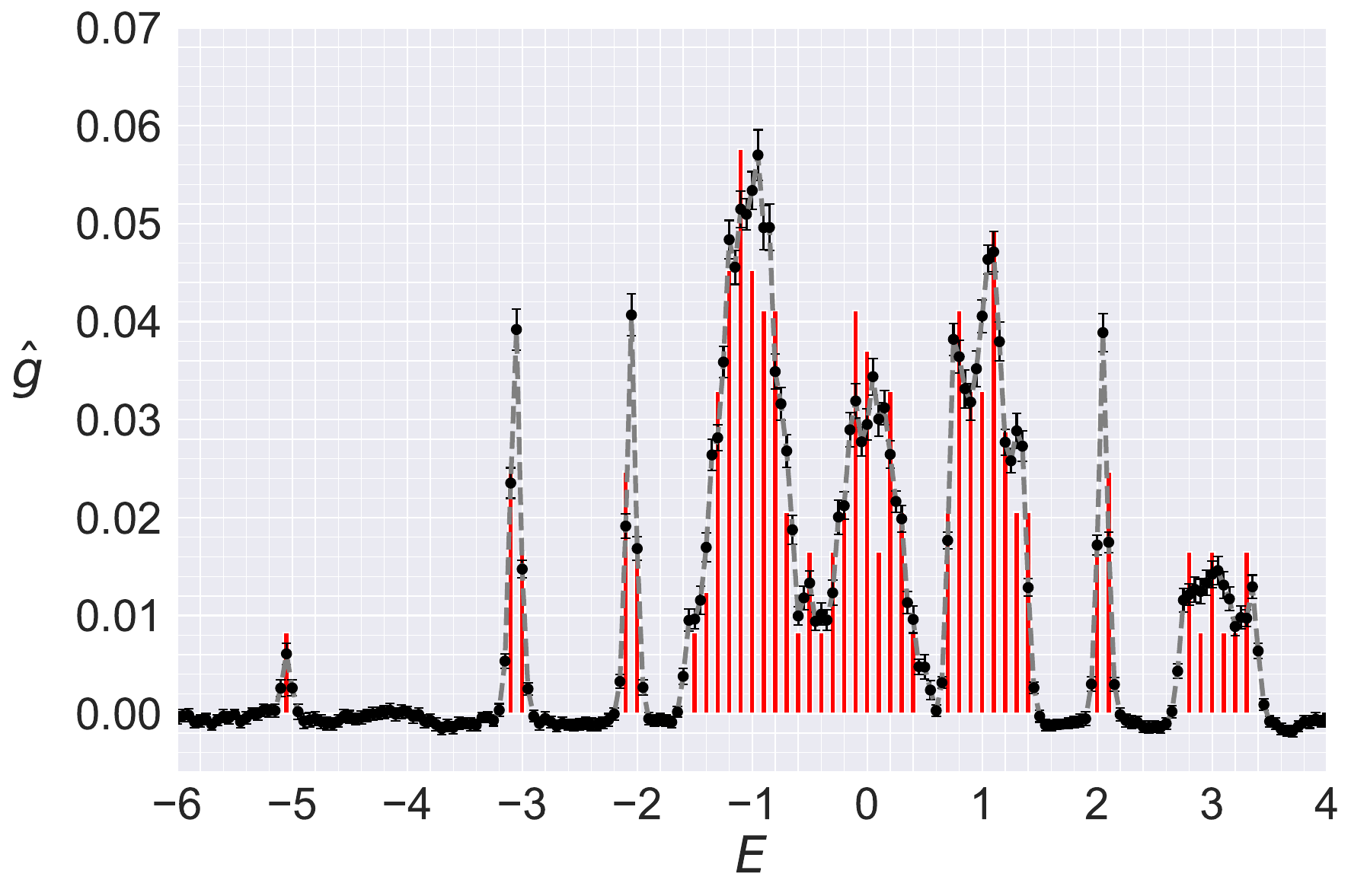} \\ (a) & (b)
\end{tabular}
\caption{Kernel-broadened DoS reconstruction for the one-dimensional spin--$1$ model using the Rodeo kernel method. For each target energy, the Rodeo algorithm was evaluated using $1\,000$ independent evolution times sampled from a Gaussian distribution with standard deviation $\sigma=20$. Panels (a) and (b) correspond to transverse-field strengths $B/(JS)=0.05$ and $0.2$, respectively.}
\label{fig:Potts}
\end{figure*}

Fig.~\ref{fig:Potts} presents the reconstructed kernel-broadened DoS for transverse magnetic fields $B/(JS)=0.05$ and $B/(JS)=0.2$, shown in panels (a) and (b), respectively.  Owing to the enhanced sensitivity of the spectrum to the magnetic field, the exact Hamiltonian eigenvalues exhibit no degeneracies larger than two. Consequently, for visualization purposes only, the exact eigenvalues were grouped into energy bins of width $0.1$, corresponding to the intervals $[E,E+0.1)$. This binning provides a clearer visual representation of the DoS by avoiding the use of color gradients to indicate the local concentration of eigenvalues. All numerical calculations, however, were performed using the original unbinned spectrum.

As discussed previously, a high concentration of eigenvalues within a narrow energy interval causes the individual Gaussian contributions to overlap, preventing the resolution of the corresponding peaks in the Rodeo response. For the spin--$1$ model considered here, the eigenvalues are so closely spaced that resolving the merged peaks by increasing the standard deviation of the Gaussian sampling distribution would require prohibitively large evolution times, thereby substantially increasing both the computational cost and the errors associated with the Suzuki--Trotter decomposition. Consequently, we estimate the total degeneracy by integrating the reconstructed DoS over the corresponding peak region.

\begingroup
\addtolength{\tabcolsep}{4pt}
\begin{table}[htbp]
\centering
\caption{Comparison between the reconstructed and exact DoS for the $N=5$ spin--$1$ ring at $B/(JS)=0.05$. 
The columns list, respectively, the peak energy, the exact degeneracy $\Omega$ of the corresponding $B=0$ energy level, 
the exact normalized DoS $g$, the normalized peak-integrated estimator $\hat g$, and the corresponding relative deviation $\Delta_r g$.}
\label{tab:potts005}
\begin{tabular}{c c c c c}
\toprule
E   & $\Omega$ & $g=\Omega/N_s$ & $\hat{g}$ & $\Delta_r g$ \\
\midrule
-5  & 2  & $\approx$0.008 & 0.01(1) & 0.26 \\
-3  & 10 & $\approx$0.041 & 0.05(1) & 0.10 \\
-2  & 10 & $\approx$0.041 & 0.04(1) & 0.05 \\
-1  & 80 & $\approx$0.329 & 0.33(1) & 0.01 \\
0   & 51 & $\approx$0.210 & 0.22(1) & 0.03 \\
1   & 60 & $\approx$0.247 & 0.23(1) & 0.06 \\
2   & 10 & $\approx$0.041 & 0.04(1) & 0.01 \\
3   & 20 & $\approx$0.082 & 0.08(1) & 0.04 \\
\bottomrule
\end{tabular}
\end{table}
\endgroup

For the weak magnetic field $B/(JS)=0.05$, the transverse field only slightly lifts the degeneracies of the $B=0$ spectrum. Therefore, the integrated DoS over each broadened peak is expected to remain close to the corresponding degeneracy of the zero-field case. The comparison presented in Table~\ref{tab:potts005} confirms this expectation. In contrast, for $B/(JS)=0.2$, the spectral splitting extends over a much broader energy range, making such an integration no longer meaningful. Instead, we assess the accuracy of the reconstructed DoS by comparing the resulting thermodynamic quantities with those obtained from exact diagonalization.

\begin{figure*}[h]
\begin{tabular}{c c}
\includegraphics[scale=0.3]{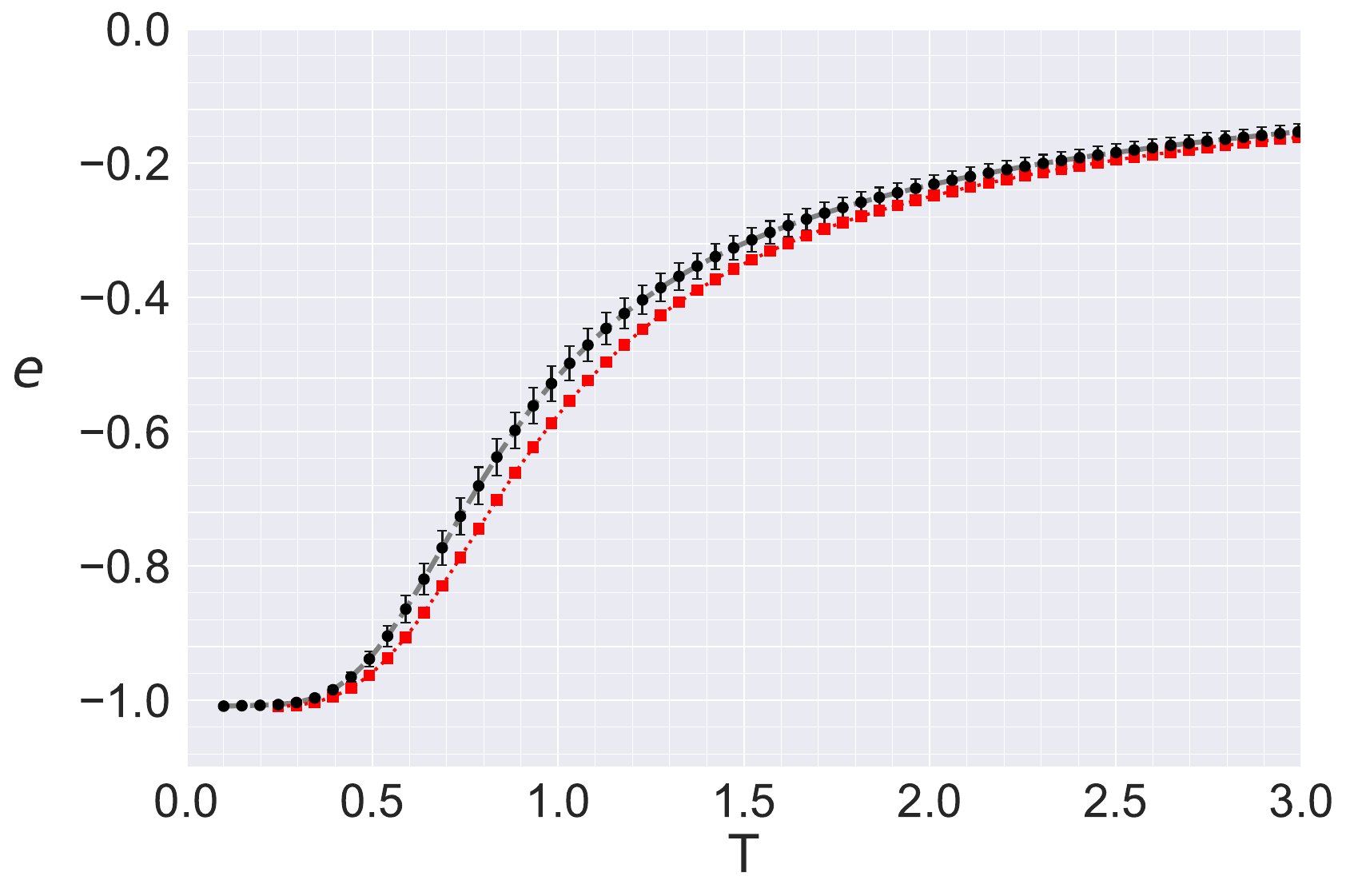}  & 
\includegraphics[scale=0.3]{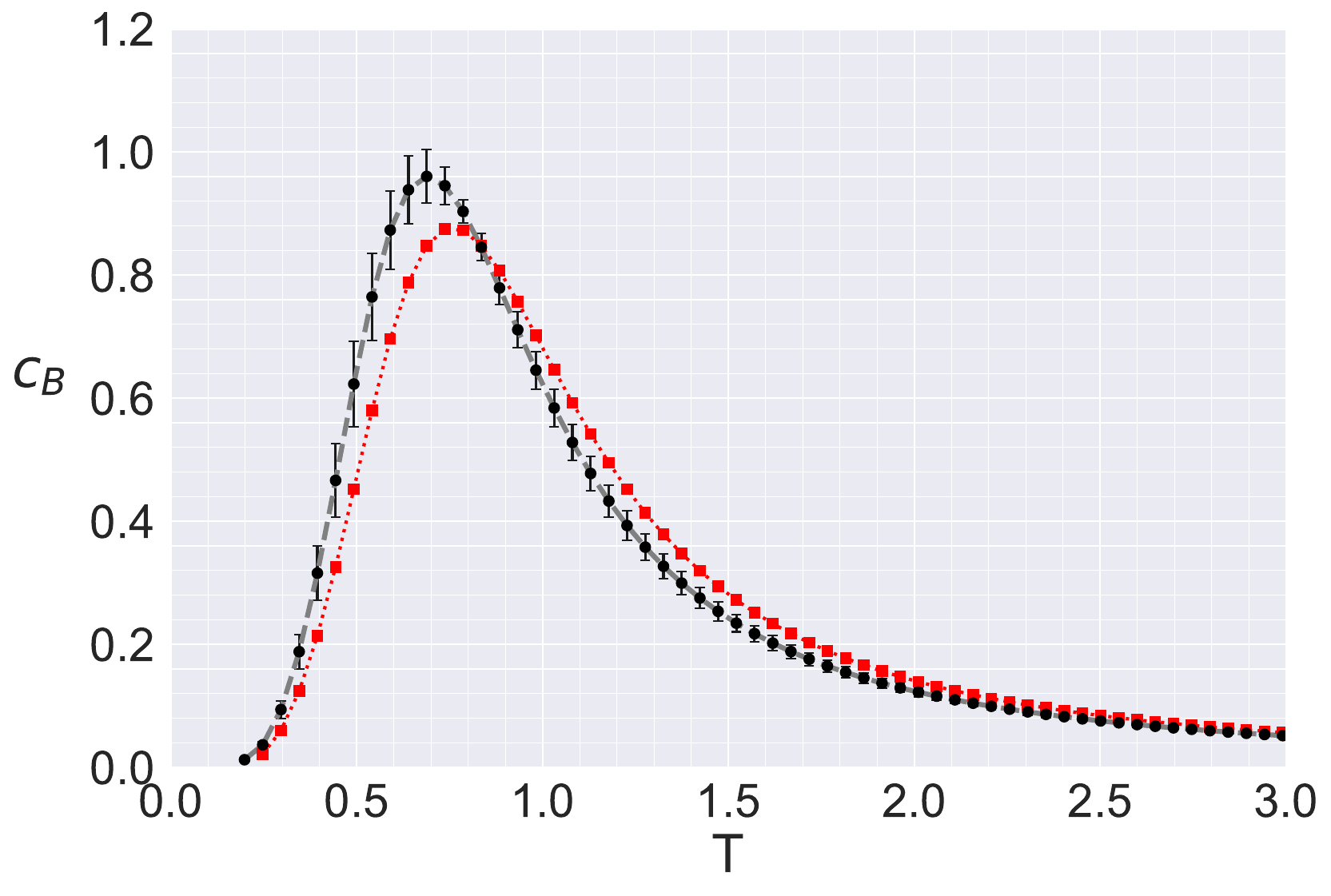} \\ (a) & (b)
\end{tabular}
\caption{Thermodynamic properties of the spin--$1$ ring with
$N=5$ spins in the presence of a transverse magnetic field $B/(JS)=0.2$.
Panels (a) and (b) show the mean energy per spin and the specific heat at constant magnetic field as functions of temperature, respectively. Black circles correspond to the results obtained using the proposed kernel-based DoS estimator, while red squares denote the
exact results obtained by direct Hamiltonian diagonalization.}
\label{fig:cB}
\end{figure*}

The canonical partition function is given by
\begin{equation}
Z(\beta)
=
\int g(E)e^{-\beta E}\,\mathrm{d}E,
\end{equation}
which is related to the Helmholtz free energy through
$F=-k_{\mathrm B}T\ln Z$. From the partition function, the mean energy per spin,
\begin{equation}
e
=
-\frac{1}{N}
\left(
\frac{\partial\ln Z}{\partial\beta}
\right)_B,
\end{equation}
and the specific heat at constant magnetic field,
\begin{equation}
c_B
=
\frac{k_B\beta^2}{N}
\left(
\frac{\partial^2\ln Z}{\partial\beta^2}
\right)_B = \frac{k_B\beta^2}{N} \Big( \langle E^2\rangle - \langle E\rangle^2 \Big),
\end{equation}
can be readily evaluated. The corresponding results are presented in Fig.~\ref{fig:cB}~(a) and (b), respectively. Here, $c_B$ is measured in units of the Boltzmann constant $k_B$.

It is worth emphasizing that, because the specific heat $c_B$ is proportional to the energy variance, it is particularly sensitive to errors in the reconstructed DoS. Since the dominant source of statistical uncertainty arises from the finite number of Haar-random input states, increasing the number of sampled states is expected to further improve the accuracy of the reconstructed thermodynamic quantities.

%%%%%%%%%%%%%%%%%%%%%%%%%%%%%%%%%%%%%%%%%%%%%%%%%%%%%%%%%%%%%%%%%%%%%%%%%%%%%%%
\section{Conclusion and Perspectives}
\label{sec:discussion}

We have shown that the Rodeo algorithm, driven by Haar-random input states,
functions as a quantum kernel method for estimating the DoS.
The estimator requires nothing beyond the standard single-ancilla Rodeo
circuit, with no bespoke state preparation other than a (pseudo)random
product of local rotations that approximates Haar weights. We
validated the construction on the one-dimensional spin-$\sfrac{1}{2}$ and spin-$1$ transverse-field Ising models: the spectral resolution is tunable through the width of the
sampling distribution, level degeneracies are recovered by integrating the
reconstructed peaks, and the thermodynamic quantities derived from the
estimated DoS agree with exact diagonalization within the statistical
uncertainties.

The broader contribution is methodological. Two mature bodies of 
knowledge---window design in classical signal processing and kernel 
damping in spectral estimation---map naturally onto the design space of 
the Rodeo algorithm, as summarized by the dictionary in Sec.~\ref{sec:dictionary}. 
Within this correspondence, the temporal sampling distribution assumes
the role of the reconstruction kernel, while the established figures of 
merit of window design---including side-lobe attenuation, main-lobe 
width, equivalent noise bandwidth, and spectral leakage---become 
concrete design parameters for tailoring the estimator to a desired 
spectral resolution or to specific hardware and decoherence constraints. 
Leveraging these well-established techniques provides a systematic route to
 optimizing the Rodeo algorithm, rather than developing analogous strategies
from first principles. The corresponding analysis of spectral leakage,
 computational cost, and decoherence will be presented in future work 
 and carries over directly to the DoS reconstruction framework developed here.

Two limitations delimit the present study and naturally motivate future work. First, exact Haar-random states are exponentially expensive to prepare. In practice, however, approximate unitary designs or random product states are often sufficient for stochastic trace estimation, as demonstrated in the classical kernel polynomial method (KPM)~\cite{weisse2006}. Quantifying the residual bias introduced by such shallow state-preparation circuits therefore constitutes a natural extension of the present work. Second, the dynamic range of the estimator is ultimately limited by the side-lobe floor of the spectral kernel and by statistical sampling noise. For applications requiring an accurate determination of $\ln g(E)$ over many decades, a promising strategy is to employ the Rodeo estimator as the energy oracle within iterative flat-histogram methods, such as Wang--Landau~\cite{Fugao} or multicanonical (MUCA)~\cite{MUCA} sampling. Indeed, if the input state is prepared such that its spectral weights satisfy $|c_j|^2\propto1/g(E_j)$, properly normalized over the spectrum, the Rodeo response becomes approximately uniform across the energy range, thereby satisfying the flat-histogram condition underlying these algorithms.

%%%%%%%%%%%%%%%%%%%%%%%%%%%%%%%%%%%%%%%%%%%%%%%%%%%%%%%%%%%%%%%
\appendix

%%%%%%%%%%%%%%%%%%%%%%%%%%%%%%%%%%%%%%%%%%%%%%%%%%%%%%%%%%%%%%%%%%%%%%%%%%%%
\section{Analysis of the Ancilla Qudit dimension effect}
\label{app:quditD}

\begin{figure}[h]
\begin{tabular}{c c}
\includegraphics[scale=0.28]{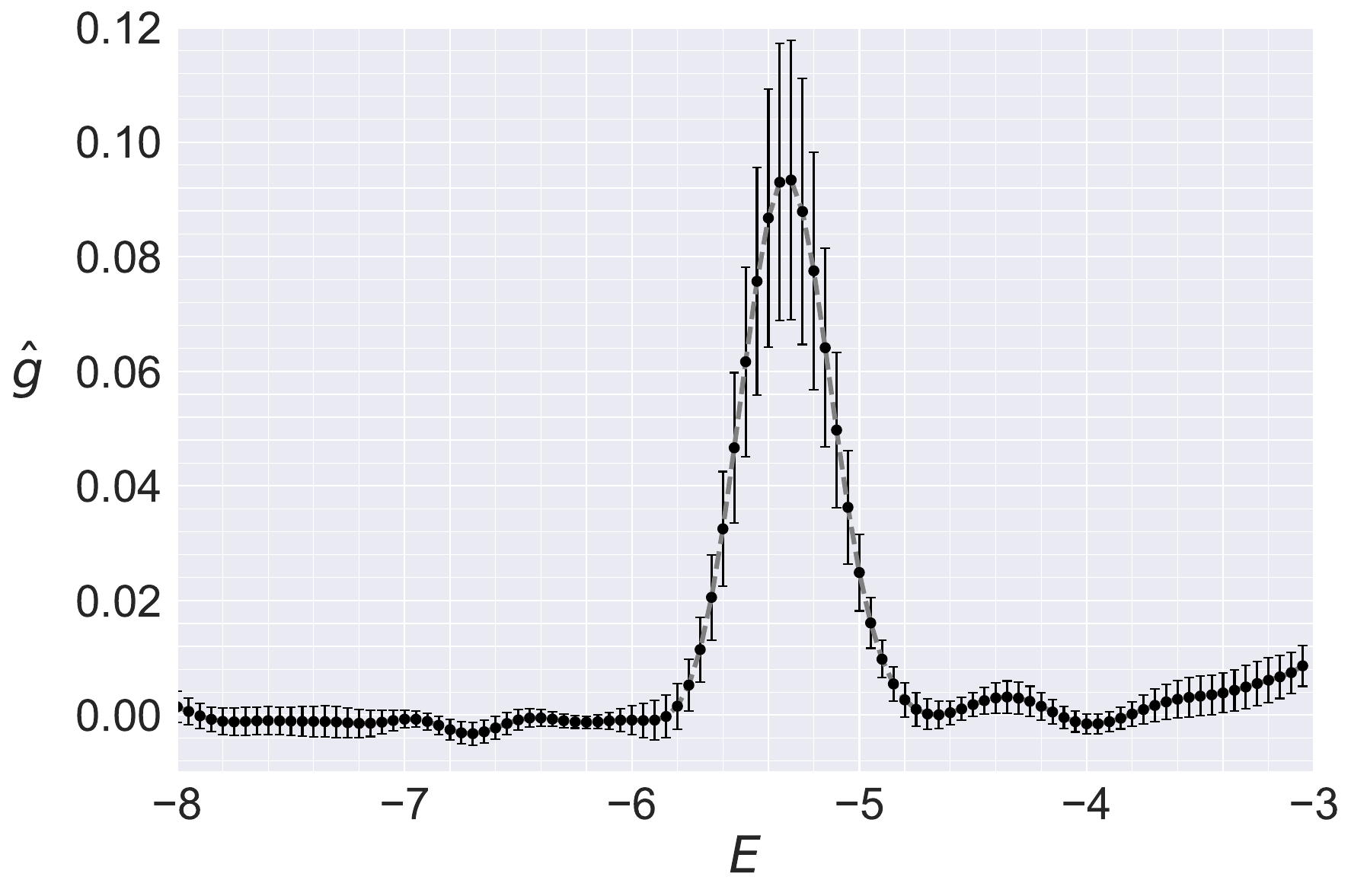} & 
\includegraphics[scale=0.28]{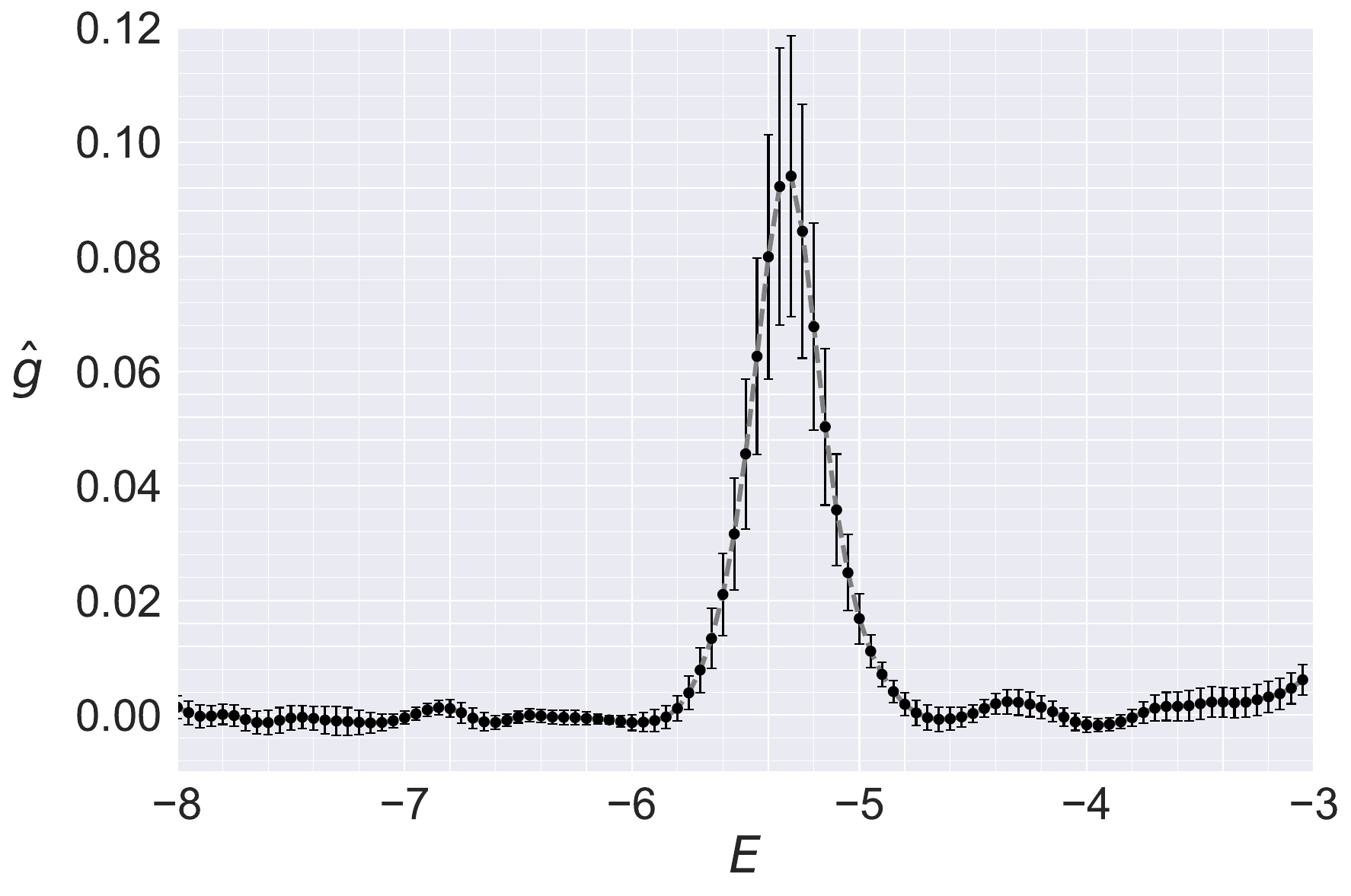} \\ (a) & (b) \\
\includegraphics[scale=0.28]{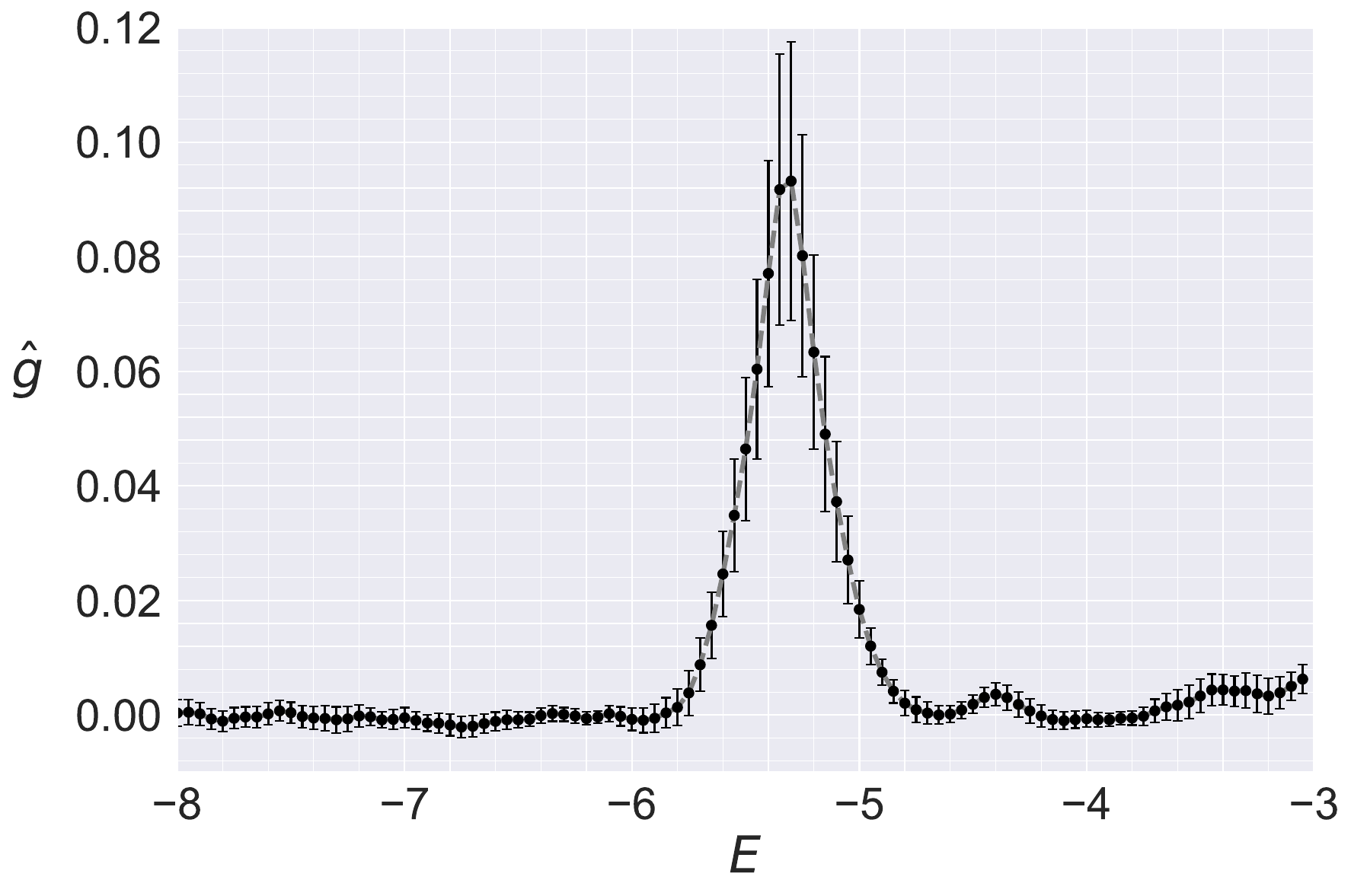} & 
\includegraphics[scale=0.28]{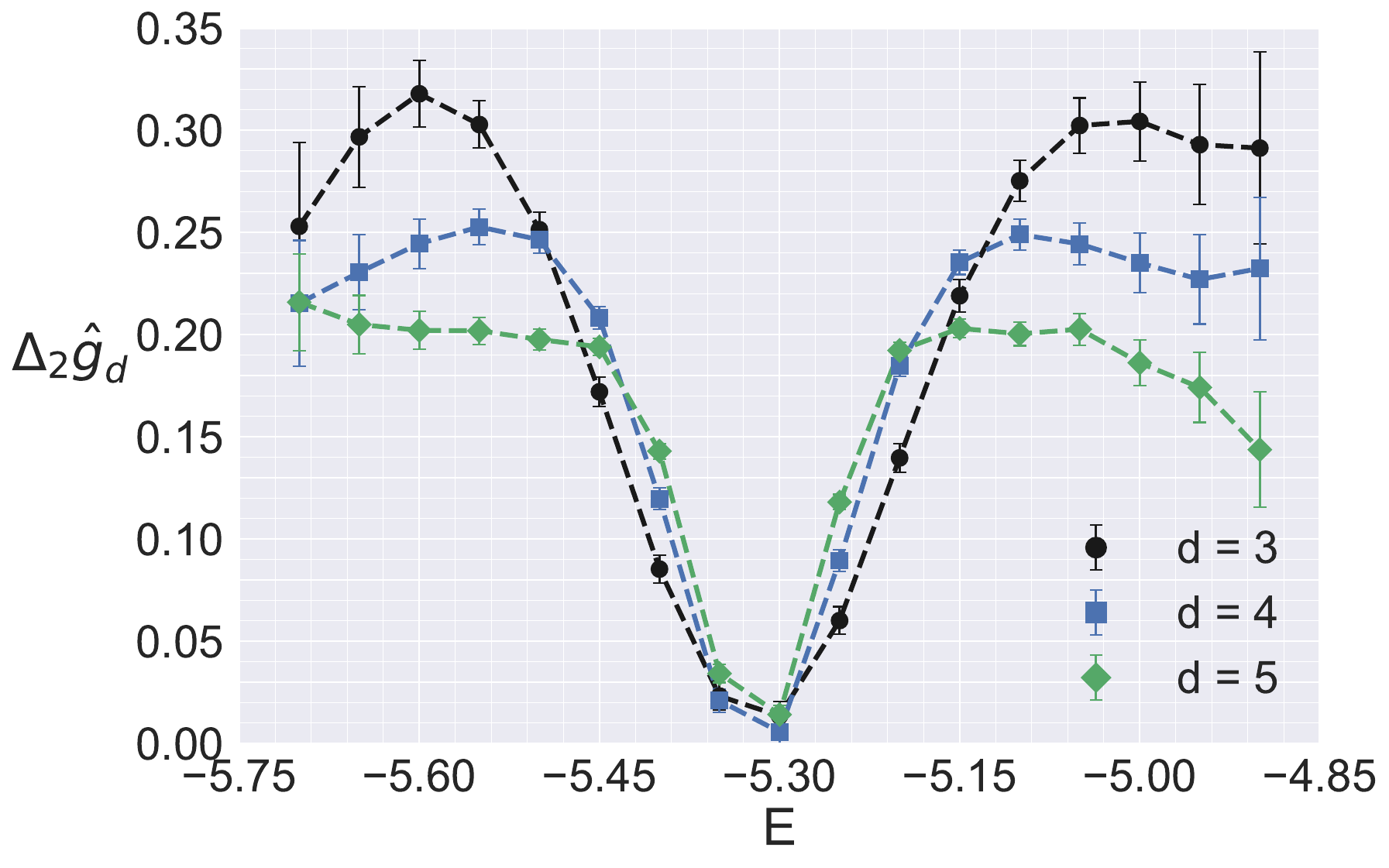} \\ (c) & (d)
\end{tabular}
\caption{Zoom at the lowest-energy peak of the kernel-broadened DoS reconstruction using the Rodeo kernel method for the one-dimensional spin-$\sfrac{1}{2}$ transverse-field Ising model with field strength $B/(JS)=0.5$. For each target energy, the Rodeo algorithm was evaluated using $1\,000$ independent evolution times sampled from a Gaussian distribution with standard deviation $\sigma=5$. Panels (a)--(c) show the results for ancilla-qudit dimensions $d=2$, $3$, and $4$, respectively. Panel (d) shows the relative differences between the results for $d=3$, $4$, and $5$ and those for $d=2$, as defined in eqn.~\eqref{eq:relDiff}. }
\label{fig:peakLowE}
\end{figure}

The effect of the ancilla-qudit dimension on spectral resolution is investigated in this appendix. Fig.~\ref{fig:peakLowE} shows the lowest-energy peak of DoS reconstruction for the one-dimensional spin-$\sfrac{1}{2}$ Ising model in the presence of a transverse field $B/(JS)=0.5$. Panels (a), (b), and (c) correspond to ancilla qudits of dimensions $d=2$, $3$, and $4$, respectively.  We draw \(1{,}000\) independent evolution times from a zero-mean Gaussian distribution with standard deviation \(\sigma = 5.0\). The estimator was averaged over \(R=10\) Haar-random initial states. The same set of initial states and evolution times was used for all values of $d$ to minimize statistical fluctuations due to the finite number of random realizations, thereby isolating the effect of the ancilla dimension. Visual inspection indicates that the peak is broader for $d=2$. Moreover, away from the peak, the fluctuations around zero and the associated error bars are larger than those observed for $d=3$ and $4$.

Following Ref.~\cite{rocha2026qudit}, we use the relative difference between the DoS reconstruction obtained using an ancilla qudit and that obtained with an ancilla qubit as a metric for the peak narrowing effect. This difference is defined as
\begin{equation}
   \Delta_2 \hat{g}_d(E) =  1 - \frac{\hat{g}_d(E)}{\hat{g}_2(E)}.
    \label{eq:relDiff}
\end{equation}
In Fig.~\ref{fig:peakLowE}~(d), we show the relative differences between the results obtained using ancilla qudits of dimensions $d=3$, $4$, and $5$ and those obtained using an ancilla qubit ($d=2$). These differences are represented by black circles, blue squares, and green diamonds, respectively. The dashed lines serve as guides to the eye, connecting symbols of the same color. Since the reconstructed DoS for $d=4$ and $d=5$ are visually similar, we omit the $d=5$ results from the corresponding plot and present only their relative difference. These results support the claim that the qutrit ($d=3$) implementation yields the most pronounced narrowing of the spectral peak.

Similarly, following Ref.~\cite{rocha2026qudit}, we quantify the fluctuations using the standard error of the $\hat{g}$ estimates far from the spectral peak. As shown in that reference, far from the eigenvalues, the standard error due to temporal sampling is
\begin{equation}
\sigma_{\mathrm{gauss}}(\mathcal{Z}) = \sqrt{\frac{1}{2N_t}}
\end{equation}
for a qubit ancilla, whereas for $d>2$ it is given by
\begin{equation}
\sigma_{\mathrm{gauss}}(\mathcal{Z}) = \frac{1}{d}\sqrt{\frac{d^2 - 2d +2}{2N_t}}.
\end{equation}
Far from the eigenvalues, $\sigma_{\mathrm{haar}}(\mathcal{R})\approx 0$, see Appendix~\ref{app:typicality}. Therefore, those expressions yield the following theoretical prediction for the standard error of $\hat{g}$:
\begin{equation}
    \sigma(\hat g)=\frac{\sigma_{\mathrm{gauss}}(\mathcal{Z})}{\sqrt{R}}.
\end{equation}

Table~\ref{tab:error} summarizes the results: the first column lists the ancilla dimension $d$, the second reports the theoretical value for $\sigma(\hat g)$, the third reports the fluctuation, and the fourth gives the average error-bar magnitude. The last column quantifies the reduction ratio in fluctuations for $d=3$, $4$, and $5$ compared with the ancilla-qubit implementation ($d=2$). These results qualitatively support the hypothesis that the ancilla-qutrit implementation ($d=3$) yields the largest reduction in fluctuations among the dimensions considered.
\begin{table}[h]
    \caption{DoS fluctuation estimates for different ancilla dimensions. The columns report the ancilla dimension \(d\), the theoretical standard deviation of the DoS far from any peak position, its empirical measurement as the fluctuation metric, the average of the error bar, and the relative fluctuation reduction with respect to the qubit-ancilla implementation.}
    \label{tab:error}
    \begin{tabular}{@{}ccccc@{}}
    \toprule
     \hspace{0.2cm}   $d$  \hspace{0.2cm}   \hspace{0.2cm}
     &\hspace{0.2cm} $\sigma(\hat g)$ \hspace{0.2cm} &  \hspace{0.2cm}Fluctuation   \hspace{0.2cm} &  Error bar \hspace{0.2cm} 
     &\hspace{0.2cm} reduction ratio \hspace{0.2cm} \\ 
     \midrule
2&$\approx 0.00708$ & $0.00250(8)$ & $0.0087(3)$ &  -- \\
3&$\approx 0.00527$ & $0.00208(4)$ & $0.0068(3)$ & $16.8\%$ \\
4&$\approx 0.00560$ & $0.00235(5)$ & $0.0069(3)$ & $6\%$ \\
5&$\approx 0.00583$ & $0.00248(5)$ & $0.0072(2)$ & $0.8\%$ \\
    \botrule
    \end{tabular}
\end{table}

The final consideration concerns the error bars at the peak position. In this region, for a low-dimensional Hilbert space, fluctuations across Haar-random input states can dominate the uncertainty. These fluctuations are expected to scale as $\mathcal{O}(1/\sqrt{RN_s})$, as shown in eqn.~(\ref{eq:sigmaHaar_g}). For the five-spin-$\sfrac{1}{2}$ Ising model, $N_s=2^5=32$. With $R=10$ Haar-random input states, this scaling gives an estimate of $1/\sqrt{RN_s}\approx 0.056$, which is larger than the error bars of approximately $0.03$ observed at the peak position in Fig.~\ref{fig:peakLowE}.

\begin{figure}[h]
\begin{tabular}{c c}
\includegraphics[scale=0.28]{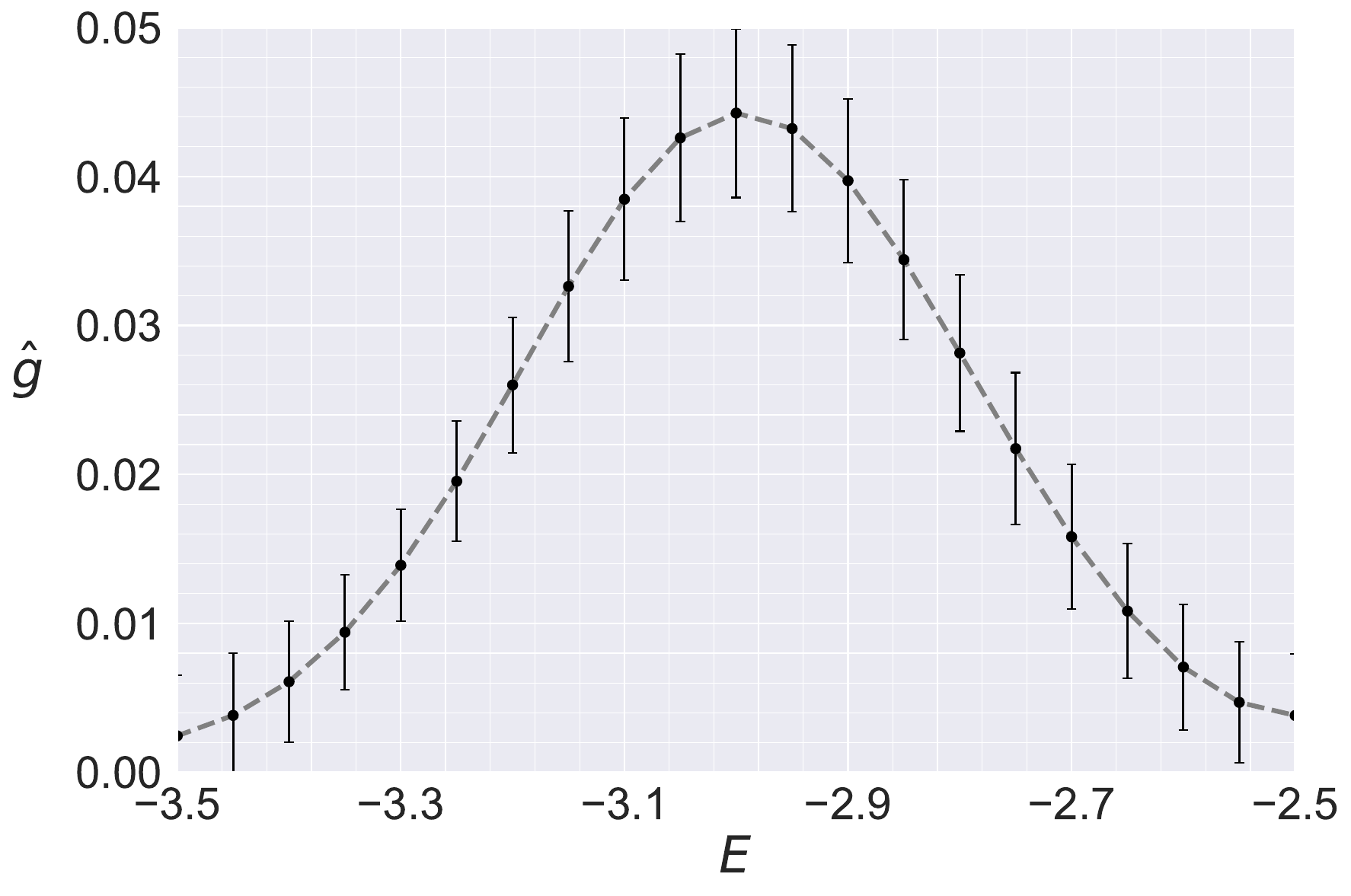} & 
\includegraphics[scale=0.28]{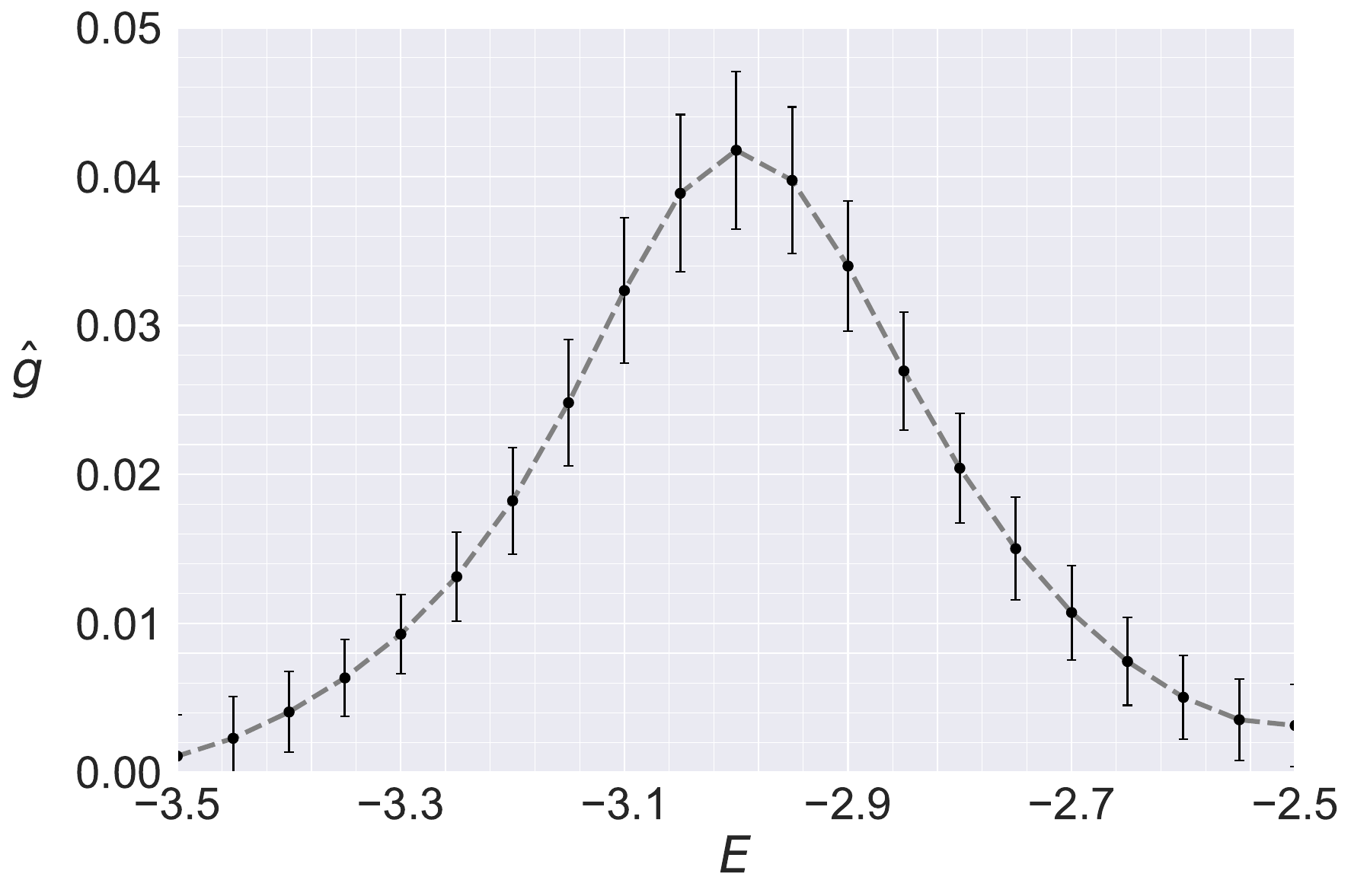} \\ (a) & (b) 
\end{tabular}
\caption{Zoom at the second lowest-energy peak of the kernel-broadened DoS reconstruction using the Rodeo kernel method for the one-dimensional spin-$1$ transverse-field Ising model with field strength $B/(JS)=0.05$. For each target energy, the Rodeo algorithm was evaluated using $1\,000$ independent evolution times sampled from a Gaussian distribution with standard deviation $\sigma=5$. Panels (a) and (b) show the results for ancilla-qudit dimensions $d=2$ and $3$, respectively. }
\label{fig:S1peakLowE}
\end{figure}

To further investigate the dependence of the peak-position uncertainty on the Hilbert-space dimension, Fig.~\ref{fig:S1peakLowE} shows the second-lowest-energy peak in the reconstructed DoS for the one-dimensional five-spin-$1$ Ising model in the presence of a transverse field $B/(JS)=0.05$. Panels (a) and (b) correspond to the ancilla qubit and ancilla qutrit, respectively. We use the same parameters as in the previous example to focus on the effect of $N_s$: $1{,}000$ independent evolution times sampled from a zero-mean Gaussian distribution with standard deviation $\sigma=5.0$, and $R=10$ Haar-random initial states. As before, the peak is narrower for the ancilla qutrit. In this case, the scaling estimate is $1/\sqrt{RN_s}\approx 0.02$, which is also larger than the error bars of approximately $0.005$ observed at the peak position. 

Overall, the observed error bars are qualitatively consistent with the predicted $\mathcal{O}(1/\sqrt{RN_s})$ scaling, although the numerical values differ from the estimates based on this scaling alone.
It is worth noting that this scaling relation does not determine the numerical prefactor, and the actual uncertainty depends on the spectral kernel and the distribution of spectral weights.

%%%%%%%%%%%%%%%%%%%%%%%%%%%%%%%%%%%%%%%%%%%%%%%%%%%%%%%%%%%%%%%%%%%%%%%%%%%%

\section{Fluctuations of the Rodeo Response for Haar-Random States}
\label{app:typicality}

In this appendix we derive the statistical fluctuations of the Rodeo response
for Haar-random initial states. The calculation relies only on the second and
fourth moments of Haar-random pure states and makes explicit the origin of the $N_s^{-1/2}$ suppression associated with quantum typicality.

%The typicality of the expectation value can be expressed as 
%\begin{equation}
%    \langle \mathcal{O}\rangle \approx \overline{\langle \mathcal{O}\rangle}
%\label{eq:defTypical}
%\end{equation}
%Where $\langle \mathcal{O}\rangle = \langle \psi| \mathcal{O} |\psi \rangle$ denotes the quantum expectation value of the operator $\mathcal{O}$ in the state $|\psi\rangle$, and $\overline{\langle \mathcal{O}\rangle}$ denotes the ensemble average of this expectation value over random states. 
%This relation follows from the properties of random states, $|\psi\rangle = \sum_k c_k|k\rangle$, distributed uniformly on the unit hypersphere, for which  $\overline{c_k} = 0$ and $\overline{c_{\ell}^* c_j} = \delta_{\ell,j}/N_s$, see ref.~\cite{Facchi2015}. Then, for large Hilbert spaces, eqn.~(\ref{eq:defTypical}) tends to equality, i.e.
%\begin{equation}
%    \lim_{N_s\to \infty} \langle \mathcal{O}\rangle = \overline{\langle \mathcal{O}\rangle}.
%\end{equation}
%Sparsely populated spectral regions of $\mathcal{O}$ contribute less to $\langle \mathcal{O}\rangle$. In a practical sense, the limit 
%Relative accuracy can remain limited in sparsely populated spectral regions.

%%%%%%%%%%%%%%%%%%%%%%%%%%%%%%%%%%%%%%%%%%%%%%%%%%%%%%%%%%%%%

\subsection{Variance}

The variance follows from

\begin{equation}
\operatorname{Var}(\mathcal{R})
=
\mathbb E(\mathcal{R}^2)
-
\mathbb E(\mathcal{R})^2,
\end{equation}
where we adopted $\mathcal{R} = \mathcal{R}_{d_a}(E,\psi)$ to simplify the notation. Expanding the square,

\begin{equation}
\mathcal{R}^2
=
\sum_{ij}
|c_i|^2
|c_j|^2
G_iG_j.
\end{equation}
where $G_i = G(E-E_i)$, with $E_i$ being the $i$-th energy eigenvalue.

The only ingredient required is the Haar fourth moment,
\begin{equation}
\mathbb E
\left[
c_i c_j^*
c_k c_l^*
\right]
=
\frac{
\delta_{ij}\delta_{kl}
+
\delta_{il}\delta_{jk}
}
{N_s(N_s+1)},
\label{eq:haar4}
\end{equation}
which immediately gives
\begin{equation}
\mathbb E
\left[
|c_i|^2|c_j|^2
\right]
=
\frac{1+\delta_{ij}}
{N_s(N_s+1)}.
\label{eq:haarprob}
\end{equation}
Hence
\begin{align}
\mathbb E(\mathcal{R}^2)
&=
\sum_{ij}
G_iG_j
\,
\frac{1+\delta_{ij}}
{N_s(N_s+1)}
\\
&=
\frac{
\left(\sum_iG_i\right)^2
+
\sum_iG_i^2
}
{N_s(N_s+1)}.
\end{align}
Using
\begin{equation}
\mathbb E(\mathcal{R})^2
=
\frac1{N_s^2}
\left(
\sum_iG_i
\right)^2,
\end{equation}
we obtain
\begin{align}
\operatorname{Var}(\mathcal{R})
&=
\frac{
\left(\sum_iG_i\right)^2
+
\sum_iG_i^2
}
{N_s(N_s+1)}
-
\frac{
\left(\sum_iG_i\right)^2
}
{N_s^2}
\\
&=
\frac{
N_s\sum_iG_i^2
-
\left(\sum_iG_i\right)^2
}
{N_s^2(N_s+1)}.
\label{eq:exactvariance}
\end{align}

Introducing the spectral kernel average
\begin{equation}
\overline G
=
\frac1{N_s}
\sum_iG_i,
\end{equation}
the numerator satisfies
\begin{equation}
N_s\sum_iG_i^2
-
\left(\sum_iG_i\right)^2
=
N_s
\sum_i
(G_i-\overline G)^2,
\end{equation}
so that
\begin{equation}
\operatorname{Var}(\mathcal{R})
=
\frac{
\displaystyle
\sum_i
(G_i-\overline G)^2
}
{N_s(N_s+1)}.
\label{eq:compactvariance}
\end{equation}
Eqn.~(\ref{eq:compactvariance}) shows that the fluctuations are
completely determined by the variance of the spectral kernel over the
eigenvalue distribution. When the probe energy $E$ is far from all eigenvalues, the kernel values $G_i$ are approximately zero, and hence $\operatorname{Var}(\mathcal{R})\approx 0$. Conversely, when $E$ lies near one or more eigenvalues, the kernel values can vary across the spectrum, giving rise to nonzero fluctuations.

%%%%%%%%%%%%%%%%%%%%%%%%%%%%%%%%%%%%%%%%%%%%%%%%%%%%%%%%%%%%%

\subsection{Typicality scaling}

Since the filter $G(\Delta)$ is bounded independently of the Hilbert-space dimension,

\begin{equation}
\sum_i
(G_i-\overline G)^2
=
O(N_s),
\end{equation}
and therefore
\begin{equation}
\operatorname{Var}(\mathcal{R})
=
O(N_s^{-1}),
\end{equation}
or equivalently,
\begin{equation}
\sigma(\mathcal{R})
=
O(N_s^{-1/2}).
\end{equation}

Thus the fluctuations of the Rodeo response decrease as the inverse square
root of the Hilbert-space dimension, which is the characteristic signature
of quantum typicality.

We emphasize that the Hilbert space dimension-dependent suppression in eqn.~(\ref{eq:compactvariance}) concerns the absolute uncertainty of $\mathcal{R}$. The relative uncertainty, however, depends on the ratio between the standard deviation and the mean of the estimator. Since $\mathbb{E}(\mathcal{R})=\overline{G}$ can be small in spectral regions where few eigenvalues lie within the kernel window, the relative uncertainty can remain large even when the absolute uncertainty is suppressed by increasing $N_s$. Consequently, the relative accuracy may be limited in sparsely populated spectral regions, such as the low-energy tail of the spectrum.
%%%%%%%%%%%%%%%%%%%%%%%%%%%%%%%%%%%%%%%%%%%%%%%%%%%%%%%%%%%%%

\subsection{Variance of the DoS estimator}

For the estimator
\begin{equation}
\hat g(E)
=
\frac1R
\sum_{r=1}^{R}
\mathcal{R}_{d_a}(E,\psi_r),
\end{equation}
constructed from $R$ independent Haar-random states, the sample mean is unbiased, $\mathbb{E}(\hat g)=\mathbb{E}(\mathcal{R})$, and its variance is
\begin{equation}
\operatorname{Var}(\hat g)
=
\frac{1}{R}\operatorname{Var}(\mathcal{R}).
\end{equation}
Consequently, at fixed $R$, the standard error is given by
\begin{equation}
\sigma(\hat g)
=
\frac{O(N_s^{-1/2})}{\sqrt{R}} = O\left(\frac{1}{\sqrt{RN_s}}\right).
\end{equation}
Since $N_s=d_s^N$, the variance decreases exponentially with the system
size and consequently
\begin{equation}
\lim_{N\rightarrow\infty}
\sigma(\hat g) = 0.
\end{equation}
Hence, in the thermodynamic limit, the fluctuations associated with the
Haar-random input states become negligible in absolute terms, and the overall uncertainty is
determined primarily by the statistical error of the Rodeo algorithm, whose
analysis is presented in Ref.~\cite{rocha2026qudit}.

%%%%%%%%%%%%%%%%%%%%%%%%%%%%%%%%%%%%%%%%%%%%%%%%%%%%%%%%%%%%%%%%%%%%%%%%%%%%
\section{Hann (raised-cosine) sampling distribution}
\label{app:hann}

An attractive alternative to the Gaussian sampling law is the Hann
(raised-cosine) distribution,
\begin{equation}
p_{\mathrm{H}}(t)=
\begin{cases}
\dfrac{1}{t_{\max}}
\cos^2\!\left(\dfrac{\pi t}{2t_{\max}}\right),
&
|t|\le t_{\max},
\\[2ex]
0,
&
\text{otherwise},
\end{cases}
\label{eq:hann}
\end{equation}
which is normalized over the interval $[-t_{\max},t_{\max}]$. Using the
trigonometric identity $\cos^2 x=\tfrac12(1+\cos 2x)$,
eqn.~(\ref{eq:hann}) can be rewritten as
\begin{equation}
p_{\mathrm{H}}(t)
=
\frac{1}{2t_{\max}}
\left[
1+
\cos\!\left(\frac{\pi t}{t_{\max}}\right)
\right],
\qquad
|t|\le t_{\max}.
\label{eq:hann2}
\end{equation}

The average of the Rodeo kernel, eqn.~(\ref{eq:Kernel}), over $p(t)$ can be seen as the 
Fourier transform of the sampling distribution. Thus, the spectral kernel can be written as
\begin{equation}
G(E)
=
\int_{-\infty}^{\infty}
p_{\mathrm{H}}(t)\,
e^{-iEt}\,dt.
\label{eq:hannFT}
\end{equation}
Because $p_{\mathrm{H}}$ is even, eqn.~(\ref{eq:hannFT}) coincides with the characteristic 
function of eqn.~(\ref{eq:charfun}), $G(E)=\Phi(E)=\Phi(-E)$, so the sign convention in the 
exponent is immaterial; for a qubit ancilla, $d_a=2$, the general filter of
eqn.~(\ref{eq:Ggeneral}) reduces precisely to this quantity.
Substituting eqn.~(\ref{eq:hann2}) yields
\begin{equation}
G(E)
=
\frac{1}{2t_{\max}}
\left(
I_1
+
I_2
\right),
\end{equation}
where
\begin{align}
I_1
&=
\int_{-t_{\max}}^{t_{\max}}
e^{-iEt}\,dt,
\\
I_2
&=
\int_{-t_{\max}}^{t_{\max}}
\cos\!\left(\frac{\pi t}{t_{\max}}\right)
e^{-iEt}\,dt.
\end{align}

The first integral is simply
\begin{equation}
I_1
=
\frac{2\sin(Et_{\max})}{E} = \frac{2Et_{\max}^2\sin(Et_{\max})}{E^2t_{\max}^2}.
\end{equation}
Writing the cosine as $\cos x=\tfrac12\left(e^{ix}+e^{-ix}\right)$,
the second integral becomes
\begin{equation}
I_2
=
\frac12
\int_{-t_{\max}}^{t_{\max}}
e^{-i(E-\pi/t_{\max})t}\,dt
+
\frac12
\int_{-t_{\max}}^{t_{\max}}
e^{-i(E+\pi/t_{\max})t}\,dt,
\end{equation}
which evaluates to
\begin{equation}
I_2
=
-
\frac{2E\sin(Et_{\max})}
{E^2-\pi^2/t_{\max}^2} = \frac{2Et_{\max}^2\sin(Et_{\max})}
{\pi^2 - E^2t_{\max}^2},
\end{equation}
where we have used the identity
$\sin(x\pm\pi)=-\sin x$.

Combining both contributions leads to the closed-form expression
\begin{equation}
G(E)
=
\frac{\pi^2
\sin(Et_{\max})}
{Et_{\max}
\left(
\pi^2-E^2t_{\max}^2
\right)}
\,.
\label{eq:hannKernel}
\end{equation}
L'Hôpital's rule guarantees the analyticity at $E=0$ and $E=\pm\pi/t_{\max}$. 
Moreover, continuity and the normalization of \(p_{\mathrm{H}}\) imply
\begin{equation}
G(0)=1
\quad  \text{and}  \quad
G\!\left(\pm\frac{\pi}{t_{\max}}\right)=\frac12.
\end{equation}
 It is worth noting that the former relation holds for any probability distribution, whereas the latter is specific to the Hann distribution.
 
 Considering an ancilla qudit of dimension $d_a$, the general spectral kernel of eqn.~(\ref{eq:Ggeneral}) yields
 %\begin{widetext}
 \begin{equation}
 %\begin{aligned}
G(E)
= %&
\frac{d_a-1}{d_a}\frac{\pi^2
\sin(Et_{\max})}
{Et_{\max}
\left(
\pi^2-E^2t_{\max}^2
\right)} + %\\
%&
\frac{1}{d_a}\frac{\pi^2
\sin\!\left((d_a-1)Et_{\max}\right)}
{(d_a-1)Et_{\max}
\left(
\pi^2-(d_a-1)^2E^2t_{\max}^2
\right)}
\,.
\label{eq:hannKernelQudit}
%\end{aligned}
\end{equation}
%\end{widetext}
%
This relation reduces to eqn.~(\ref{eq:hannKernel}) in the limit $d_a\rightarrow\infty$, where the 
finite-dimensional correction vanishes. As discussed in Ref.~\cite{rocha2026qudit}, finite
 values of $d_a$ introduce an additional interference term in the reconstruction kernel. 
 This correction is most pronounced for the qutrit implementation ($d_a=3$) and rapidly 
 decreases with increasing ancilla dimension, becoming negligible in the large-$d_a$ limit.

Moreover, introducing the dimensionless variable $x=Et_{\max}$,
eqn.~(\ref{eq:hannKernel}) assumes the compact form
\begin{equation}
\tilde G(x)
=
\frac{\pi^2\sin x}
{x(\pi^2-x^2)}
=
\mathrm{sinc}(x)
+
\frac12\,\mathrm{sinc}(x-\pi)
+
\frac12\,\mathrm{sinc}(x+\pi),
\label{eq:hannCompact}
\end{equation}
where $\mathrm{sinc}(x)=\sin(x)/x$. The second equality, obtained by
recombining the three terms over a common denominator, makes explicit
that the Hann kernel is the linear combination of three shifted sinc
functions anticipated in Table~\ref{tab:windows}.

For large energies, the kernel behaves as
\begin{equation}
G(E)
\simeq
-\frac{\pi^2}{t_{\max}^3}\,
\frac{\sin(Et_{\max})}{E^3},
\qquad
|E|\rightarrow\infty,
\end{equation}
so the side-lobe envelope decays as
$|G(E)|\le \pi^2/\!\left(t_{\max}^3|E|^3\right)$.
The Hann sampling distribution therefore suppresses spectral leakage far
more efficiently than the uniform distribution, whose sinc kernel decays
only as $|E|^{-1}$, while retaining the experimentally convenient finite
temporal support that the Gaussian law lacks.
%%%%%%%%%%%%%%%%%%%%%%%%%%%%%%%%%%%%%%%%%%%%%%%%%%%%%%%%%%%%%%%%%%%%%%%%%%%%
\section{Data set, Averages, and Error Analysis}
\label{app:error}
In our calculations, we construct a data set containing the probe energy, $E$, the evolution time sampled from the Gaussian distribution, $t$, and the ancilla-qubit expectation value, $\mathcal{Z}=\mathcal{Z}(E,t,\psi_r)$, defined in eqn.~(\ref{eq:H_echo}). Each data file corresponds to one Haar-random initial state, $|\psi_r\rangle$. Thus, we generate $R$ data files, each containing $N_t$ entries for every probe energy $E$.

For each initial state, the $n$-th raw moment of $\mathcal{Z}$ with respect to the sampled evolution times is calculated as
\begin{equation}
    \langle\mathcal{Z}^n\rangle_t
    =\frac{1}{N_t}\sum_{j=1}^{N_t}
    \mathcal{Z}^n(E,t_j,\psi_r).
    \label{eq:SAnPratical}
\end{equation}
The first raw moment gives the spectral amplitude (SA),
\begin{equation}
\mathcal{R}_{d_a}(E,\psi_r)=\langle\mathcal{Z}\rangle_t.
\label{eq:SAPratical}
\end{equation}
This is an estimate of the Gaussian average in eqn.~(\ref{eq:mixture}). It is worth emphasizing that sampling evolution times from the Gaussian distribution allows the corresponding Gaussian-weighted average to be estimated by an arithmetic mean.

Assuming independent evolution-time samples, the standard deviation of the estimated SA is therefore
\begin{equation}
    \sigma(\mathcal{Z})
    =\sqrt{\frac{
    \langle\mathcal{Z}^2\rangle_t
    -\langle\mathcal{Z}\rangle_t^2
    }{N_t-1}}.
    \label{eq:sigmaGaussVar}
\end{equation}
If the DoS is estimated using a single initial state, we obtain $\hat{g}(E)=\langle\mathcal{Z}\rangle_t\pm\sigma(\mathcal{Z})$. 
 Otherwise, we must account for the statistical fluctuations across the ensemble of Haar-random initial states by considering the (n)-th raw moments of the SA:
\begin{equation}
    \langle\mathcal{R}^n\rangle_r
    =\frac{1}{R}\sum_{r=1}^{R}
    \mathcal{R}_{d_a}^n(E,\psi_r).
    \label{eq:rawHarr}
\end{equation}

As defined in eqn.~(\ref{eq:ghat}), the first moment gives the reconstructed DoS,
\begin{equation}
    \hat{g}(E)=\langle\mathcal{R}\rangle_r.
\end{equation}
Using eqns.~(\ref{eq:SAnPratical}) and (\ref{eq:SAPratical}) it leads to
\begin{align}
    \hat{g}(E)&=\langle\langle\mathcal{Z}\rangle_t\rangle_r,\\
              & =\frac{1}{RN_t}\sum_{r=1}^{R}\sum_{j=1}^{N_t}
    \mathcal{Z}(E,t_j,\psi_r)
\end{align}
The standard error associated with averaging over the $R$ initial states is
\begin{equation}
    \sigma(\hat g)=\frac{s_{\mathcal{R}}}{\sqrt{R}},
\end{equation}
where 
\begin{equation}
    s^2_{\mathcal{R}}
    = \frac{ \sum_{r=1}^R\left(\mathcal{R}_{d_a}(E,\psi_r) - \hat g\right)^2
    }{R-1}.
\end{equation}
The reconstructed DoS is consequently reported as
\begin{equation}
    \hat{g}(E) = \langle\langle\mathcal{Z}\rangle_t\rangle_r \pm\sigma(\hat{g}).
\end{equation}

%%%%%%%%%%%%%%%%%%%%%%%%%%%%%%%%%%%%%%%%%%%%%%%%%%%%%%%%%%%%%%%%%%%%%%%%%%%%%%%
\section*{Acknowledgments}
The author would like to acknowledge helpful conversations with Dr. Rodrigo A. Dias.

The author acknowledges the use of large language models (Claude by Anthropic and ChatGPT by OpenAI) during the preparation of this manuscript. These tools assisted in improving the scientific writing, refining the mathematical exposition, discussing alternative theoretical formulations, and editing the \LaTeX{} manuscript. All scientific ideas, derivations, numerical simulations, and conclusions were conceived, validated, and approved by the author.

This research received no specific grant from any funding agency in the public, commercial, or not-for-profit sectors.

%%%%%%%%%%%%%%%%%%%%%%%%%%%%%%%%%%%%%%%%%%%%%%%%%%%%%%%%%%%%%%%

\section*{Conflict of Interest}
The author has no conflicts of interest to disclose.

%%%%%%%%%%%%%%%%%%%%%%%%%%%%%%%%%%%%%%%%%%%%%%%%%%%%%%%%%%%%%%%

\section*{Data Availability}
The numerical code and raw data that support the findings of this study are openly available in the repository of Ref.~\cite{data_zenodo}.

\bibliography{quantumKernelMethod}

\end{document}